\documentclass[nofootinbib,pre,superscriptaddress,twocolumn,longbibliography,floatfix]{revtex4-1}
\pdfoutput=1
\usepackage[utf8]{inputenc}
\usepackage{hyperref}
\usepackage{graphicx}  
\usepackage{dcolumn}   
\usepackage{bm}        
\usepackage{amssymb,amsmath}   
\usepackage{uri}
\usepackage{xr}

\usepackage[x11names, rgb, html]{xcolor}
\definecolor{sbase03}{HTML}{002B36}
\definecolor{sbase02}{HTML}{073642}
\definecolor{sbase01}{HTML}{586E75}
\definecolor{sbase00}{HTML}{657B83}
\definecolor{sbase0}{HTML}{839496}
\definecolor{sbase1}{HTML}{93A1A1}
\definecolor{sbase2}{HTML}{EEE8D5}
\definecolor{sbase3}{HTML}{FDF6E3}
\definecolor{syellow}{HTML}{B58900}
\definecolor{sorange}{HTML}{CB4B16}
\definecolor{sred}{HTML}{DC322F}
\definecolor{smagenta}{HTML}{D33682}
\definecolor{sviolet}{HTML}{6C71C4}
\definecolor{sblue}{HTML}{268BD2}
\definecolor{scyan}{HTML}{2AA198}
\definecolor{sgreen}{HTML}{32CD32}
\hypersetup{
  colorlinks = true,
  citecolor = {sgreen},
  urlcolor = {sblue}
}

\usepackage{footnote}
\usepackage{footmisc}

\usepackage{xcolor}
\definecolor{orange1}{rgb}{1, .467, 0}
\definecolor{blue1}{rgb}{.36, .494, .714}

\newcommand{\beginsupplement}{%
 \setcounter{table}{0}
  \renewcommand{\thetable}{S\arabic{table}}%
  \setcounter{figure}{0}
  \renewcommand{\thefigure}{S\arabic{figure}}%
  \setcounter{equation}{0}
  \renewcommand{\theequation}{S\arabic{equation}}%
}

\renewcommand{\bm}[1]{\boldsymbol{#1}}

\begin{document}
\title{Measuring irreversibility from learned representations of biological patterns}
\author{Junang Li}
\thanks{These authors contributed equally to this work.}
\affiliation{Department of Physics, Massachusetts Institute of Technology, Cambridge, MA 02139, USA}
\affiliation{Center for the Physics of Biological Function, Princeton, NJ, 08540, USA}
\author{Chih-Wei Joshua Liu}
\thanks{These authors contributed equally to this work.}
\affiliation{Department of Physics, Massachusetts Institute of Technology, Cambridge, MA 02139, USA}
\author{Michal Szurek}
\affiliation{Department of Physics, Massachusetts Institute of Technology, Cambridge, MA 02139, USA}
\author{Nikta Fakhri}
\email[Corresponding author: ]{fakhri@mit.edu}
\affiliation{Department of Physics, Massachusetts Institute of Technology, Cambridge, MA 02139, USA}

\begin{abstract}
Thermodynamic irreversibility is a crucial property of living matter. 
Irreversible processes maintain spatiotemporally complex structures and functions characteristic of living systems. 
In high-dimensional biological dynamics, robust and general quantification of irreversibility remains a challenging task due to experimental noise and nonlinear interactions coupling many degrees of freedom.
Here we use deep learning to identify tractable, low-dimensional representations of phase-field patterns in a canonical protein signaling process --- the Rho-GTPase system --- as well as complex Ginzburg-Landau dynamics.
We show that factorizing variational autoencoder neural networks learn informative pattern features robustly to noise.
Resulting neural-network representations reveal signatures of mesoscopic broken detailed balance and time-reversal asymmetry in Rho-GTPase and complex Ginzburg-Landau wave dynamics.
Applying the compression-based Ziv-Merhav estimator of irreversibility to representations, we recover irreversibility trends across complex Ginzburg-Landau patterns varying widely in spatiotemporal frequency and noise level.
Irreversibility estimates from representations similarly recapitulate cell-activity trends in a Rho-GTPase patterning system undergoing metabolic inhibition.
Additionally, we find that our irreversibility estimates serve as a dynamical order parameter, distinguishing stable and chaotic dynamics in these nonlinear systems.
Our framework leverages advances in deep learning to offer robust, model-free measurements of nonequilibrium and nonlinear behavior in complex living processes.
\end{abstract}

\maketitle

\section{Introduction}
Living matter consumes free energy through metabolism, forming patterns in processes such as development and motility ~\cite{needleman2017active,gnesotto2018broken,marchetti2013hydrodynamics,seifert2012stochastic,murugan2016biological}.
These nonequilibrium processes violate governing principles of equilibrium systems, such as the Boltzmann distribution, impeding physical characterization~\cite{crooks1999entropy}.
These processes are nevertheless constrained by the second law of thermodynamics, as free-energy consumption measurably increases the entropy of the environment and accompanies broken detailed balance ~\cite{luposchainsky2013entropy,mabillard2023heat}.
Broken detailed balance entails asymmetric transition rates between pairs of microstates, a time-reversal asymmetry enabling cycles in the phase space~\cite{battle2016broken}.
This asymmetry is also described as the ``thermodynamic arrow of time'': concretely, the forward flow of events is distinguishable from its reverse ~\cite{seif2021machine}.
The statistical distinguishability of time-forward and time-reversed processes in fact quantifies thermodynamic irreversibility~\cite{parrondo2009entropy}.
Quantification of thermodynamic irreversibility is emerging as an important source of insight into nonequilibrium processes in biological and condensed-matter physics~\cite{li2019quantifying,tan2022odd,tan2021scale,gingrich2016dissipation,seifert2019stochastic}.

Irreversibility is measured as the Kullback-Leibler divergence (KLD) from the distribution of time-forward processes to the distribution of time-reversed processes, which requires sampling over many possible steady-state configurations.
In practice, KLD estimates are constrained by the limited timescales of experimental data and the complex interactions between many components and high dimensionality intrinsic to living systems.
Reliable irreversibility estimates thus normally consider only a readily observed subset of degrees of freedom, such as time-resolved trajectories of probe particles~\cite{tan2021scale}.  
However, recent deep-learning methods manipulate and synthesize complex data with relative ease, overcoming the curse of dimensionality inherent in statistical physics~\cite{bahri2020statistical,lusch2018deep,falk2021learning,schmitt2023zyxin,hernandez2023low}.
Neural networks can reduce high-dimensional signals to low-dimensional representations, potentially facilitating irreversibility quantification in complex living processes with many degrees of freedom.

Here we present a new framework based on disentangling variational neural networks to represent complex living processes as low-dimensional dynamics in a tractable latent feature space~\cite{kim2018disentangling}.
As proof of principle, we investigate nonequilibrium biochemical waves formed by Rho-GTPase signalling protein in the actomyosin cortex of the \textit{Patiria miniata} (bat sea star) oocyte~\cite{tan2020topological}.
Using deep-learned feature-space representations, we recapitulate underlying irreversibility trends in both simulated and experimental Rho patterns.
This suggests our framework provides a physically motivated indicator of activity in living systems. 
Moreover, our thermodynamic irreversibility estimates not only correctly rank the energetics of different patterns, but also serve as an order parameter indicating different dynamical regimes of this nonlinear system.

\begin{figure*}[t!]
    \centering
	\includegraphics[width=1\textwidth]{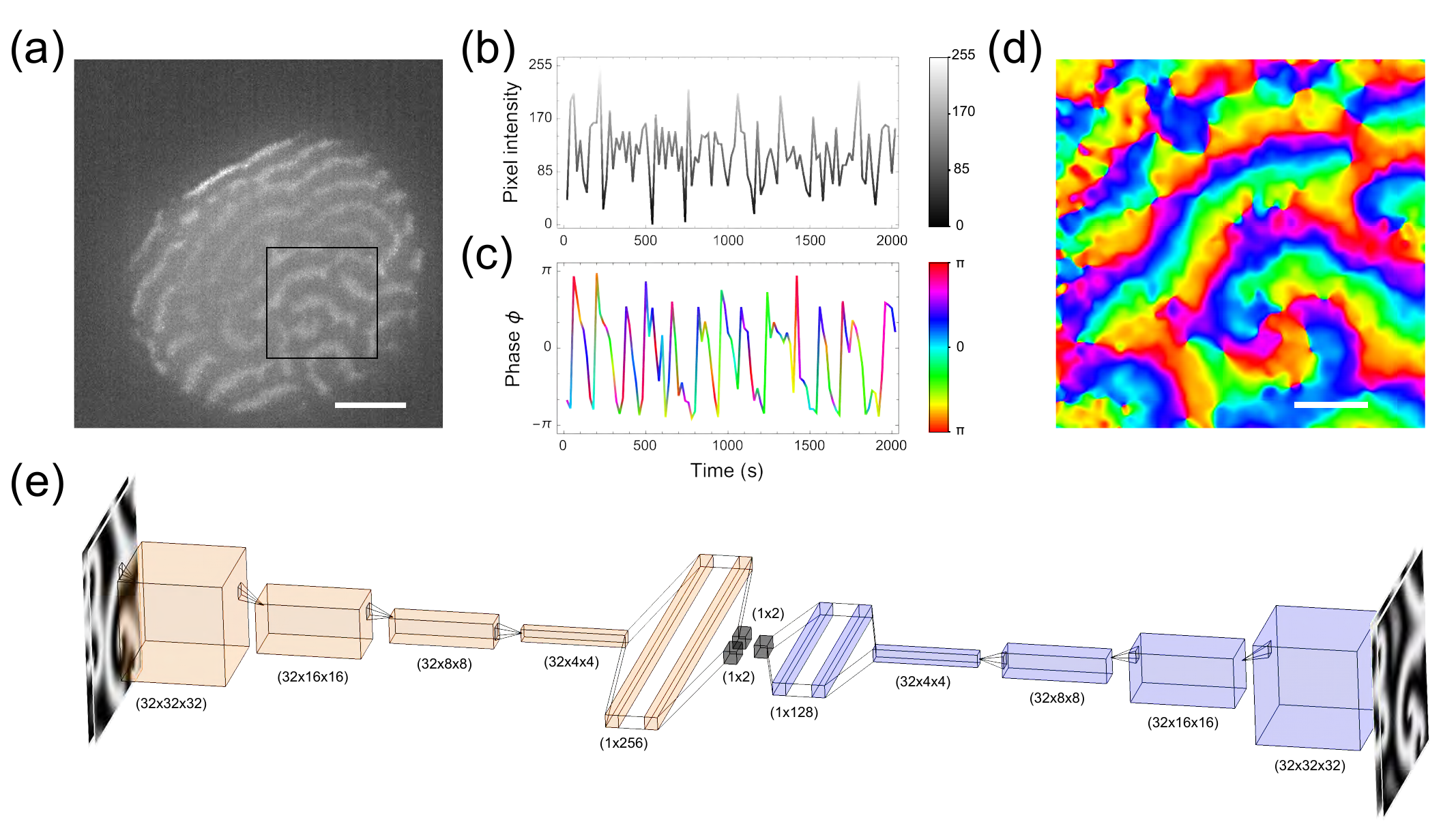}
    \caption{The FVAE architecture represents Rho patterns.
    (a) Fluorescence micrograph shows GFP-labeled Rho-GTP forming spiral waves in the membrane of a starfish oocyte. Scale bar denotes $50 \mu m$. 
    (b) A pixel in (a) oscillates noisily over 2000 seconds. Intensity oscillations show fluctuating period and amplitude. 
    (c) Relative-phase retrieval yields a clearly oscillatory signal from the intensity oscillation shown in (b).
    (d) Phase field of the boxed region in (a) retrieved by obtaining relative phases from all pixels as in (c). Scale bar denotes $15 \mu m$.
    (e) The FVAE architecture (excluding discriminator, see Methods) represents and reconstructs inputs. The encoder (orange) feeds inputs through four convolutional layers and two linear layers to the bottleneck layers (grey). The first bottleneck layer represents variational posteriors as two $1 \times 2$ vectors encoding the means and variances of two latent dimensions. The second bottleneck layer encodes $1 \times 2$ vectors sampled from the variational posterior using reparameterization, which feed into the two linear layers and four transposed convolutional layers of the decoder (blue) to reconstruct inputs. $2 \times 64 \times 64$ FVAE inputs and outputs are transformed phase-field frames.
	}
	\label{fig:schematic}
\end{figure*}

\section{Complex Ginzburg-Landau dynamics describe the experimental Rho phase field}
Evolutionarily conserved Rho GTPases play major roles in eukaryotic development~\cite{wigbers2021hierarchy}.
Membrane-associated Rho self-organizes into waves of activation, with a range of nonequilibrium steady states visualizable using fluorescent reporters specific to active, GTP-bound Rho [Fig.~\ref{fig:schematic}(a) and Methods]~\cite{tan2020topological}.
Because Rho hydrolyzes GTP and diffuses down concentration gradients as it activates and inactivates, its reaction-diffusion wave patterning consumes chemical energy and is irreversible.
Previous work indirectly inferred irreversibility in the Rho-regulated dynamics of sea-star oocytes using a subset of degrees of freedom~\cite{tan2021scale}. 
Here we seek to directly quantify irreversibility from all information encoded in fluorescently labeled Rho.

To extract dynamics from noisy experimental data, we first converted Rho intensity fields captured through fluorescence microscopy into corresponding phase fields~\cite{tan2020topological}.
Rho activation alternating with inactivation results in intensity oscillations at each pixel [Fig.~\ref{fig:schematic}(b)], from which we retrieved relative phases [Fig.~\ref{fig:schematic}(c) and Methods].   
Oscillations are more readily observed in phase fields than in intensity fields, which suffer fluctuations and envelope decay due to photobleaching and camera noise.
For example, by repeating phase retrieval for all pixels in the boxed region of Fig.~\ref{fig:schematic}(a), we generated the phase-field frame $\bm{\phi}$ in Fig.~\ref{fig:schematic}(d).

Phase dynamics of membrane Rho are captured by the complex Ginzburg-Landau (CGL) equation~\cite{tan2020topological}
\begin{equation}
	\partial_{t} A = A + (1+ic_1) \Delta A - (1+ic_2)|A|^2 A.
	\label{eq:CGLE}
\end{equation}
with $A = |A|\exp{(i\phi)}$, where $\phi$ is a phase field varying in space and time, $c_1$ models the linear dispersion of the medium, and $c_2$ models the nonlinear dispersion.
The CGL equation approximates envelope dynamics of reaction-diffusion patterning as arises in the well-known Brusselator model~\cite{falasco2018information,kuramoto1984chemical}.
Intuitively, higher $c_1$ corresponds to faster Rho diffusion, while higher $c_2$ corresponds to higher Rho activation rate~\cite{liu2021topological}.

\begin{figure}[b!]
  \centering
  \includegraphics[width=1\linewidth]{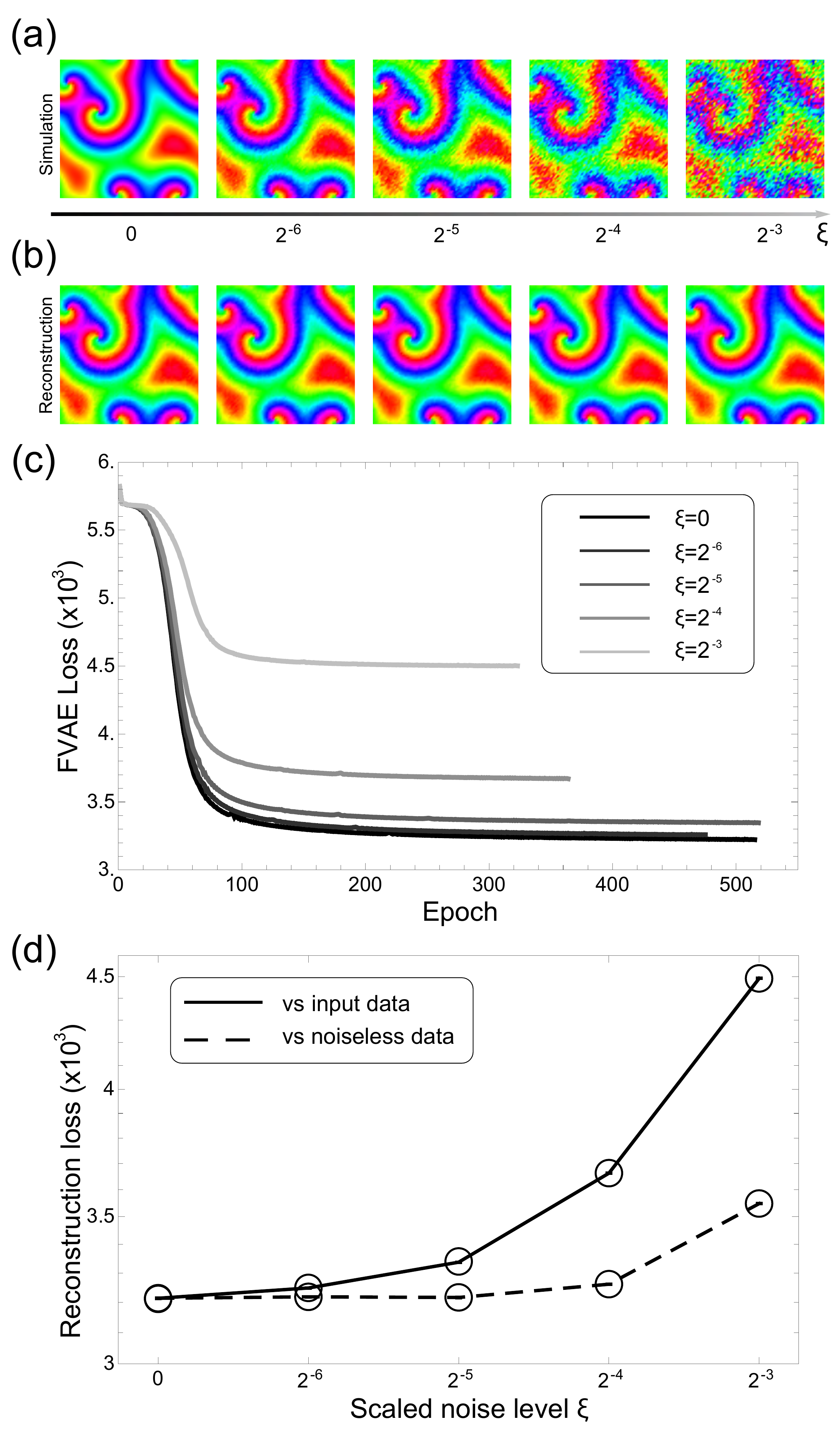}
  \caption{Trained FVAE models retrieve informative pattern features.
    (a) Representative simulated CGL phase-field frames with $c_1=-0.2$ and $c_2=0.5$ and added mean-zero Gaussian noise of varying standard deviation $2\pi\xi$.
    (b) FVAE are trained on datasets transformed from CGL phase patterns at each noise level. Depicted reconstructions of each noisy simulated phase-field frame in (a) resemble the noiseless simulated phase-field frame at $\xi =0$, regardless of the noise level of the FVAE training set.
    (c) Training loss (Eq.~\ref{eq:objective}) decreases approximately monotonically with epochs. We terminate training when loss plateaus.
    (d) FVAE trained on datasets of varying noise reconstruct their training data. Average binary cross-entropy loss measured between an FVAE reconstruction and a target is higher when the target is the noisy input sample being reconstructed than when the target is the corresponding noiseless sample, indicating that models retrieve informative, non-noise features of inputs. Error bars indicate standard errors of means over all input samples.
    }
  \label{fig:noise_result}
\end{figure}

\begin{figure*}[t!]
  \centering
  \includegraphics[width=1\linewidth]{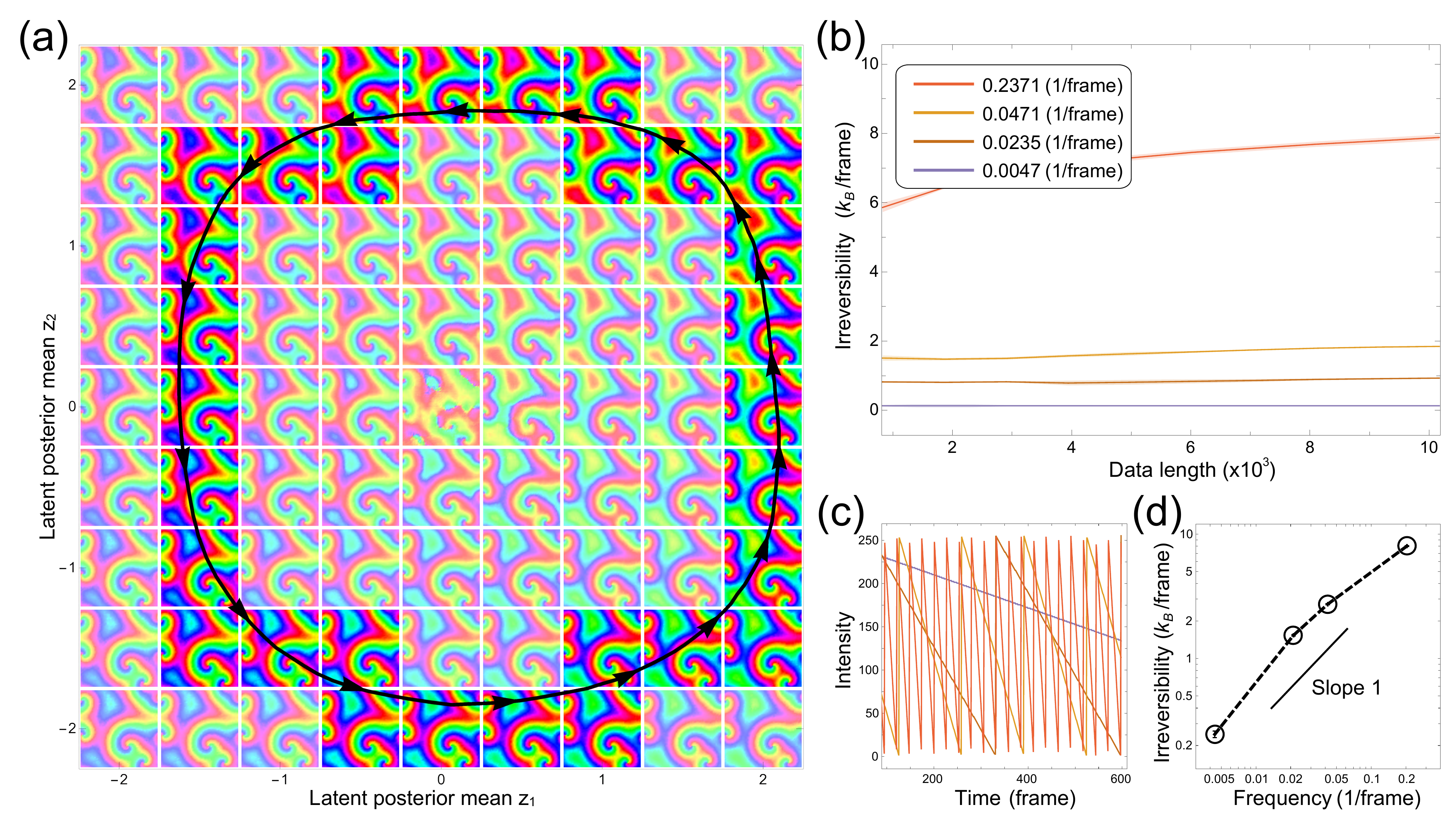}
  \caption{Latent trajectories enable irreversibility estimates.
    (a)  The trajectory (black curve) of the ($\xi =0$) pattern with snapshot shown in Fig.~\ref{fig:noise_result}(a) traces a cycle in the latent space of an FVAE trained on all of the pattern's transformed frames. Arrows indicate direction and are evenly spaced in time. Background images are phase-field frames inverse-transformed from their decoded latent-space locations. Latent-space locations visited by the trajectory show highlighted reconstructions of the pattern's phase-field frames. 
    (b) ZM irreversibility estimates computed from simulation latent trajectories increase with timestep size (oscillation frequency). Shadowed regions indicate standard errors of means calculated over three independent simulations with the same dispersions.
    (c) Pixel-oscillation time series of simulations at the four timestep sizes shown in (b).
    (d) ZM estimates of simulations increase approximately linearly with oscillation frequencies shown in (b). Error bars indicate standard errors of means calculated over three independent simulations with the same dispersions.
    }
  \label{fig:temporal_result}
\end{figure*}

\section{Factorizing variational autoencoders represent high-dimensional dynamics in a low-dimensional latent space}
Spatiotemporally continuous Rho phase fields, as in the CGL model, have many degrees of freedom and are challenging inputs for irreversibility estimators that take low-dimensional trajectories.
Crucially, we use variational autoencoders (VAE)~\cite{kingma2013autoencoding} to represent Rho and CGL phase fields in low-dimensional latent spaces.
Each VAE consists of an encoder, two bottleneck layers, and a decoder [Fig.~\ref{fig:schematic}(e) and Methods].
The encoder feeds inputs $\bm{x}$ (here transformed from $64\times 64$ phase-field frames $\bm{\phi}$) through convolutional layers followed by linear layers.
Encoder outputs in the first bottleneck layer consist of means and variances of Gaussian variational posteriors $q_{\bm{\theta}}(\bm{z}|\bm{x})$ over the $d$-dimensional latent space, where $\bm{\theta}$ denotes encoder parameters and $\bm{z}$ denotes a latent-space vector.
Decoder inputs in the second bottleneck layer are sampled from $q_{\bm{\theta}}(\bm{z}|\bm{x})$ through reparameterization, and feed through linear layers followed by transposed convolutional layers.
The decoder $p_{\bm{\psi}}(\bm{x}|\bm{z})$ outputs reconstructed $\bm{x}$, with $\bm{\psi}$ denoting decoder parameters.

Due to the periodicity of angles, phase fields $\phi$ must be transformed into neural-network training data $x$ by complex-exponentiating into two channels $(\cos\phi,\sin\phi)$ before rescaling to the range $[0,1]$ (Methods)~\cite{guyon1991design, heffernan2017capturing}.
Reconstructed phase-field pattern frames $\bm{\phi}$ are in turn inverse transformations of reconstructed VAE inputs $\bm{x}$.
For $N$ inputs $\bm{x}$, we use a factorizing VAE (FVAE) loss function~\cite{kim2018disentangling}
\begin{equation}
    \begin{aligned}
        \frac{1}{N}\sum_{i=1}^{N}\Bigl\{-\mathbb{E}_{q(\bm{z}|\bm{x}^{(i)})}[\ln{p(\bm{x}^{(i)}|\bm{z})}]\\
        +KL[q(\bm{z}|\bm{x}^{(i)})||\mathcal{N}(0,I)]\Bigl\}\\
        +KL[q(\bm{z})||\prod_{j=1}^{d}q(z_j)]
        \label{eq:objective}
    \end{aligned}    
\end{equation}
with three terms.
Here $KL[\cdot||\cdot]$ denotes the KLD, $\mathcal{N}(0,I)$ denotes a normal prior with $I$ the $d$-dimensional identity matrix, $q(\bm{z})=\frac{1}{N}\sum_{i=1}^N q(\bm{z}|\bm{x}^{(i)})$ denotes the aggregate posterior, and $q(z_j)$ denotes the aggregate-posterior marginal of $z_j$.
The first term is a binary cross-entropy reconstruction loss and measures the fidelity of FVAE reconstructions, while the second term is a regularizer for penalizing model complexity. 
The third term penalizes dependence between latent dimensions and encourages efficient (disentangled) representations (Methods).
Stochastic gradient-based optimization minimizes the loss in Eq.~\ref{eq:objective}.
Main results use $d=2$ and training batches including all transformed frames of a phase-field video [Methods and Fig.~\ref{fig:d_choice} in Supplemental Material (SM)].

By encoding high-dimensional inputs as variational posteriors in low-dimensional latent spaces, VAE can discover dynamical coordinates in nonlinear systems~\cite{gabbard2021bayesian, miles2021machine, takeishi2021physics, wang2021flow}.
To confirm that FVAE capture informative features of pattern dynamics, we trained models on CGL datasets with simulated measurement noise.
For $\xi$ varying from $0$ to $2^{-3}$, independent mean-zero Gaussians of standard deviation $2\pi\xi$ corrupt the pixels of each phase-field frame $\bm{\phi}$ for a CGL simulation with $c_1=-0.2$ and $c_2=0.5$ [representative snapshots shown in Fig.~\ref{fig:noise_result}(a)].
FVAE models train to reconstruct datasets transformed from the simulation at each noise level, with reconstructed phase-field frames shown in Fig.~\ref{fig:noise_result}(b) for the sample phase-field frames in \ref{fig:noise_result}(a).
Due to the temporal periodicity of CGL patterns, randomly sampled training and validation sets are highly similar.
As a result, we terminate training when regression over a window of epochs indicates that loss has ceased decreasing [Fig.~\ref{fig:noise_result}(c) and Methods].
Counterintuitively, though models for noisy datasets are never exposed to noiseless datasets during training, average binary cross-entropy loss between FVAE reconstructions of training data and targets is higher when the targets are the noisy input samples being reconstructed than when the targets are the corresponding noiseless samples~\ref{fig:noise_result}(d).
All reconstructed phase-field frames in Fig.~\ref{fig:noise_result}(b) closely resemble the noiseless leftmost inset of \ref{fig:noise_result}(a), consistent with previous observations that VAE denoise inputs to identify important pattern features~\cite{im2017denoising, liu2020unsupervised}.
Unless otherwise specified, we thus focus on noiseless data in subsequent simulation analyses.

\begin{figure*}[ht!]
  \centering
  \includegraphics[width=1\linewidth]{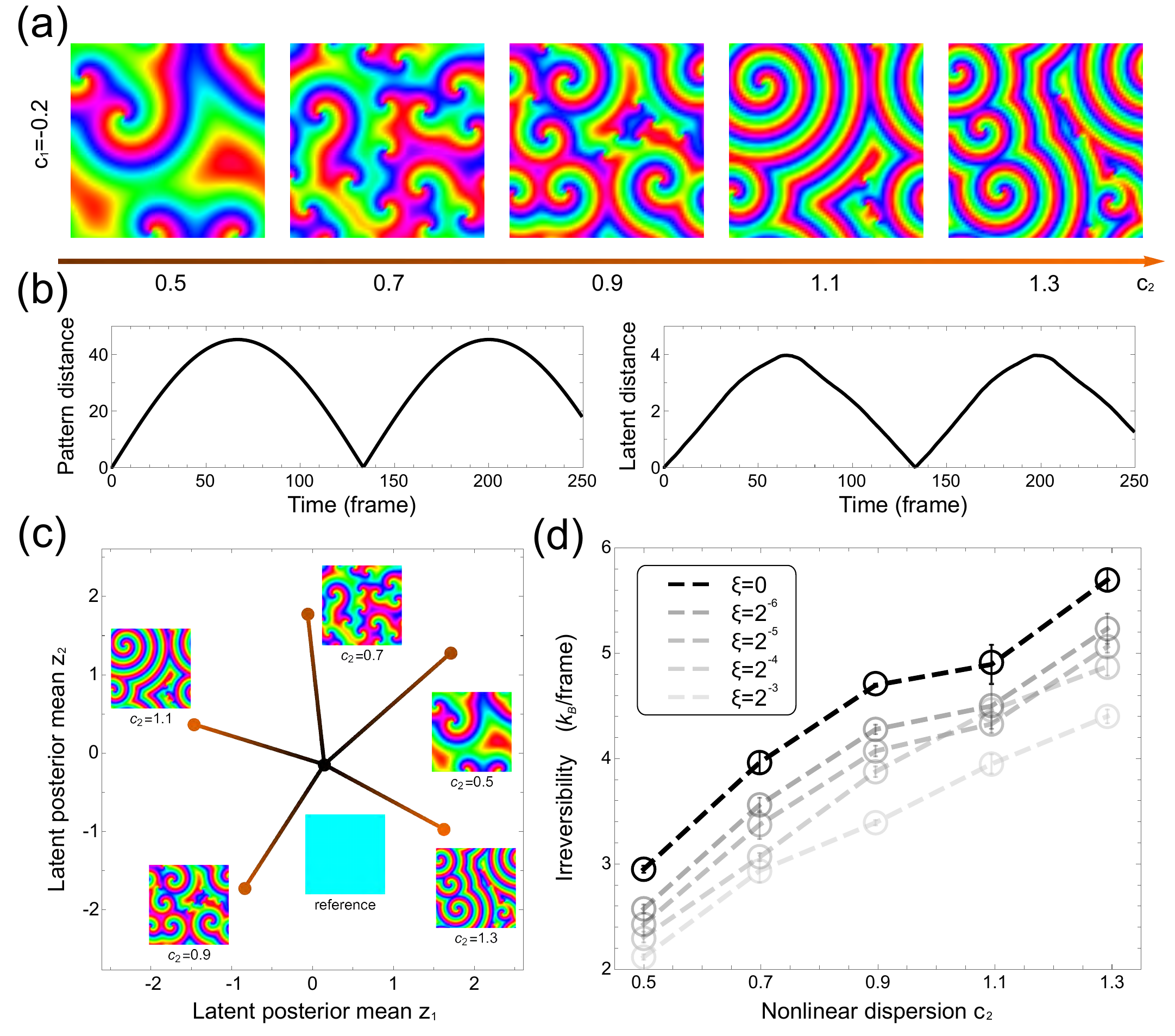}
  \caption{Rescaled latent representations enable irreversibility comparison between patterns differing in spatial structure.
    (a) Patterns with higher $c_2$ exhibit higher spatial frequencies. Snapshots show phase-field frames of patterns with linear dispersion $c_1=-0.2$ fixed and nonlinear dispersion $c_2$ varying in increments of $0.2$ between $0.5$ and $1.3$.
    (b) Distances in pattern space are approximately proportional to corresponding distances in latent spaces of models trained on patterns. Left panel shows the L2 norm between the transformed first and successive frames of a noiseless simulation with $c_1=-0.2$ and $c_2=0.5$ [Fig.~\ref{fig:noise_result}(a)]; right panel shows the L2 norm between variational-posterior means of those inputs.
    (c) Distances are shown without rescaling between representations of five representative frames in (a) and the vanishing-field reference in the superimposed latent spaces of FVAE trained on each simulation. Latent trajectories are rescaled to fix the ratio of L2 distances to the reference in input and representation spaces.
    (d) Irreversibility estimates increase with nonlinear dispersion $c_2$ and pattern complexity at fixed linear dispersion $c_1$ regardless of noise level. Means and error bars indicate averages and standard errors over three independent simulation replicates.
	}
  \label{fig:spatial_result}
\end{figure*}

\section{Latent representations enable irreversibility estimates}
As the noiseless pattern in Fig.~\ref{fig:noise_result}(a) evolves periodically, its variational-posterior mean exhibits cycles in the latent space [Fig.~\ref{fig:temporal_result}(a)].
Tiles in Fig.~\ref{fig:temporal_result}(a) are phase-field frames inverse-transformed from decoded lattice points in the latent space, with frames inverse-transformed from reconstructed inputs highlighted along the pattern-evolution trajectory.
Arrowheads denote points evenly spaced in time along a period of the latent trajectory.
FVAE latent dimensions are optimized for disentanglement, and each row or column illustrates the effect of changing one latent variable, keeping the other fixed.

Previous work demonstrated that undirectional cycles, such as that seen in the latent space, are a signature of broken detailed balance and irreversible dynamics in the context of mesoscopic biological systems~\cite{battle2016broken}.
Points along latent trajectories index pattern dynamical states captured by the FVAE neural network, evolving rapidly when the pattern evolves rapidly.
We can thus estimate pattern irreversibility by applying the Ziv-Merhav (ZM) compression estimator~\cite{ziv1977universal} of KLD rates to forward and temporally reversed coarse-grained latent trajectories (Methods, Appendix, and Fig.~\ref{fig:traj_coarse} in SM)~\cite{roldan2010estimating, roldan2012entropy}.
Resulting ZM estimates are robust to choice of tuned FVAE hyperparameters (Fig.~\ref{fig:hyperparameters_vary} in SM).
The FVAE reconstructs with loss and latent trajectories do not encode full input information.
Irreversibility estimates computed from latent trajectories are thus lower bounds.

Estimating irreversibility from latent trajectories is data-efficient and computationally fast, as illustrated in Fig.~\ref{fig:temporal_result}(b).
Note that CGL dynamics are deterministic, which results in irreversibility estimates that diverge logarithmically with increasing data length.
However, relative divergence rates are different, enabling comparisons between differently evolving patterns (Appendix).

To test our framework, we simulated CGL patterns with the same dispersion parameters but sampled at different timestep sizes [Fig.~\ref{fig:temporal_result}(c) and Methods].
Larger sampling timestep effectively increases pattern evolution speed and oscillation frequency.
Intuitively, irreversibility should increase with oscillation frequency.
The approximately linear increase in estimated irreversibility with frequency, shown in Fig.~\ref{fig:temporal_result}(d), suggests our framework successfully detects altered temporal structure and correctly orders nonequilibrium steady states by activity level.    

\section{Irreversibility estimates capture spatial scale and complexity}
Patterns differing by more than temporal frequency present additional challenges in irreversibility comparisons.
The CGL model forms patterns with diverse spatial structures, such as those in Fig.~\ref{fig:spatial_result}(a).
Spatial frequency increases with nonlinear dispersion $c_2$ in the CGL equation and with activity (effective kinetics) in the Rho system~\cite{wigbers2021hierarchy}.
Observing that reconstruction losses vary little with CGL dispersion in models trained on wave simulations (Fig.~\ref{fig:loss_phase_diagram} in SM), we adapted our framework to compare irreversibilities of CGL dynamics and Rho patterns that differ in spatial structure.
 
Consider the latent trajectory in Fig.~\ref{fig:temporal_result}(a).
Positions in FVAE latent space evolve with the encoded pattern.
Similar observations across trajectories suggest that latent-space distance scales with pattern-space distance: the L2 distance between a phase-field pattern's first transformed frame and successive transformed frames is approximately proportional to the corresponding L2 distance between the first transformed frame's variational-posterior mean and successive transformed frames' variational-posterior means [Fig.~\ref{fig:spatial_result}(b)].
In agreement with the Johnson-Lindenstrauss lemma~\cite{johnson1984extensions}, relative L2 distances are preserved between inputs $\bm{x}$ and their latent representations $\bm{z}$.
VAE obey a Lipschitz property~\cite{jordan2021provable, camuto2022variational}
\begin{equation}
	||\bm{z} - \bm{z}'|| \leq C ||\bm{x} - \bm{x}'||
    \label{eq:lipschitz}
\end{equation}
where $\bm{z}$ and $\bm{z}'$ are two latent vectors, $\bm{x}$ and $\bm{x}'$ are the corresponding transformed phase-field frames, $C>0$ is a real constant, and $|| \cdot ||$ denotes the L2 norm.
The L2 norm is thus a tractable metric for comparing scales of transformed-pattern and latent spaces.

\begin{figure*}[h!]
    \centering
        \includegraphics[width=1\linewidth]{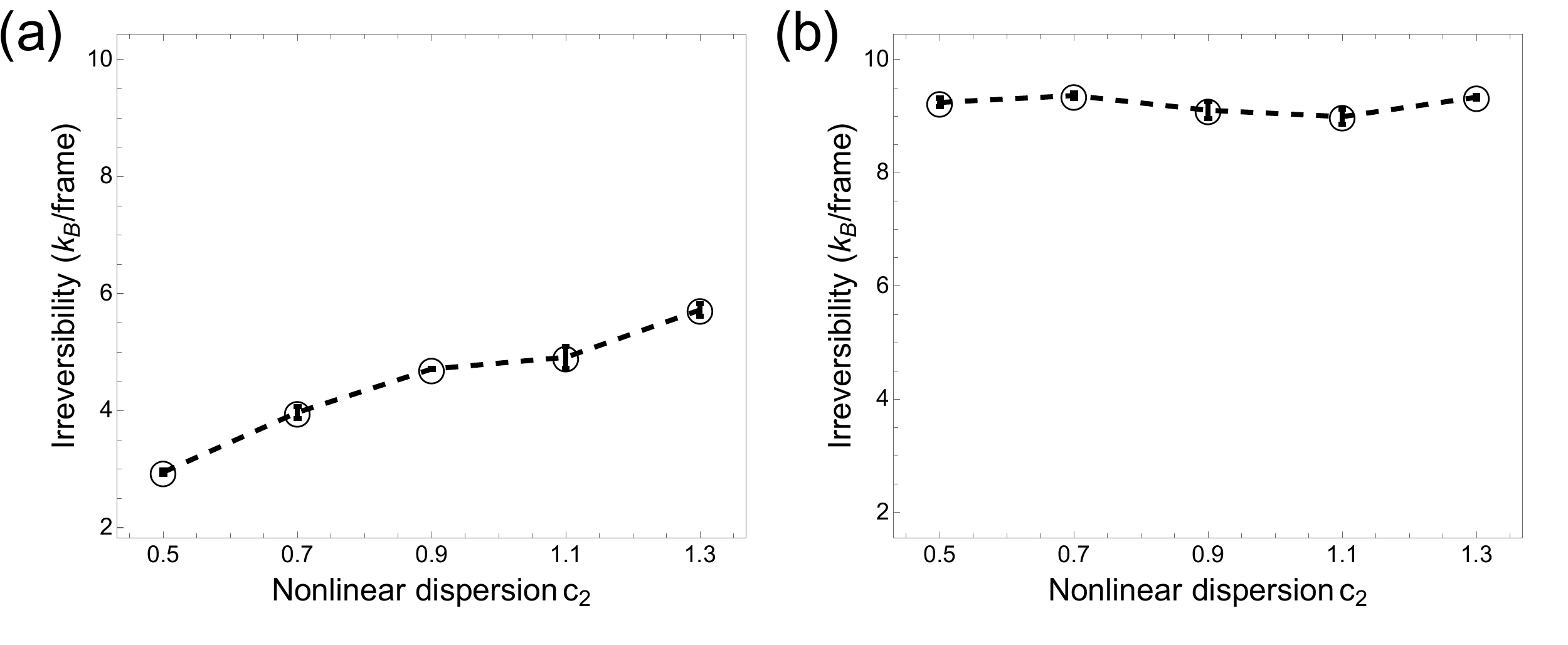}
        \caption{Local irreversibility estimates do not detect pattern irreversibility increasing with $c_2$. CGL nonlinear dispersion $c_2$ models Rho-GTPase activity level. Error bars indicate standard errors of averages over three independent simulations. (a) As shown in Fig.~\ref{fig:spatial_result}(d), our framework correctly ranks irreversibilities of CGL patterns varying in $c_2$ at fixed linear dispersion $c_1$. (b) Patterns of the same linear dispersion show the same oscillation frequency. Spatially averaged local irreversibility estimates~\cite{ro2022model} thus vary little by $c_2$ among the patterns in (a).}
        \label{fig:benchmark}
\end{figure*}

The ZM estimator requires coarse-graining of state spaces (Appendix and Fig.~\ref{fig:traj_coarse} in SM), hindering irreversibility comparisons between patterns mapped to latent spaces that differ by scaling.
Accordingly, we add a transformed phase-zero (vanishing-field) frame to each training set as a reference: the reference is mapped close to the origin in latent spaces.
Comparing L2 distances to the reference in transformed-pattern and latent spaces [Fig.~\ref{fig:spatial_result}(c)], we rescale latent trajectories to a constant distance ratio shared between models trained on different patterns (Methods).
Applying the ZM estimator to rescaled trajectories, irreversibilities increase with $c_2$ at fixed $c_1$ robustly to noise level. 
Results in Fig.~\ref{fig:spatial_result}(d) corroborate the notion that nonequilibrium potentials increase with complexity in patterns excited from homogeneous media~\cite{falasco2018information}, as well as with the interpretation of $c_2$ as modeling Rho-pathway activity level.
As increasing $c_2$ does not increase pattern evolution speed, all oscillating phase-field pixels have the same frequency in CGL simulations sharing the same $c_1$.
Unlike our framework, recently introduced local entropy production measurements that apply the ZM estimator separately to each pixel of a high-dimensional pattern thus do not detect irreversibility increasing with pattern complexity (Fig.~\ref{fig:benchmark} in SM)~\cite{ro2022model}.

\section{Irreversibility estimates rank biological states by activity}
\begin{figure}[h!]
    \centering
    \includegraphics[width=1\linewidth]{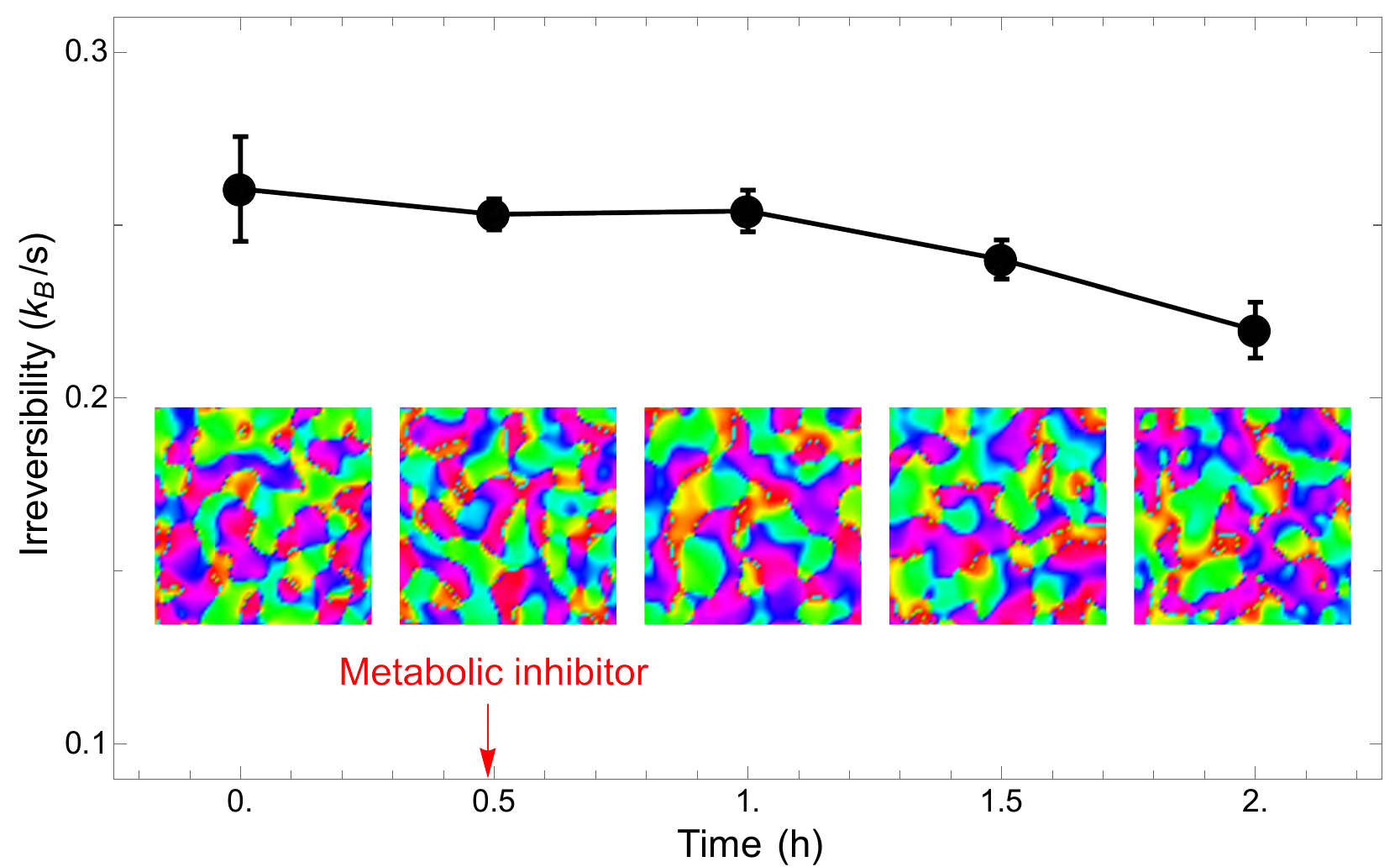}
    \caption{Irreversibility estimates recover decreased cell-activity level. Insets are representative snapshots from each non-overlapping half-hour of Rho phase-field video. Arrow shows time of oocyte treatment with sodium azide. Error bars indicate standard errors of averages over training seeds (Methods).}
    \label{fig:azide_results}
\end{figure}

Observing that our framework correctly ranks irreversibilities of simulated CGL dynamics with robustness to noise, we assessed its applicability to experimental biological data.
An oocyte forming steady-state Rho-GTPase waves was visualized for half an hour before treatment with the metabolic inhibitor sodium azide~\cite{pelling2004local}.
Sodium azide decreased cell-activity level, altering Rho-GTPase patterning over a further two hours of visualization~\cite{liu2021topological}.
We converted fluorescence-microscopy Rho-GTP intensity fields to phase fields, trained FVAE models on datasets for each half hour of phase-field video, rescaled model latent trajectories to a fixed distance ratio, and applied the ZM estimator to rescaled latent trajectories.
Resulting irreversibility estimates decrease at timepoints following treatment with sodium azide, recapitulating underlying decreases in cell-activity level (Fig. ~\ref{fig:azide_results}).
Our framework correctly ranks both simulated CGL and experimental Rho patterns by irreversibility, enabling comparisons between states with unknown relative activity levels.

\begin{figure*}[htb!]
  \centering
  \includegraphics[width=1\linewidth]{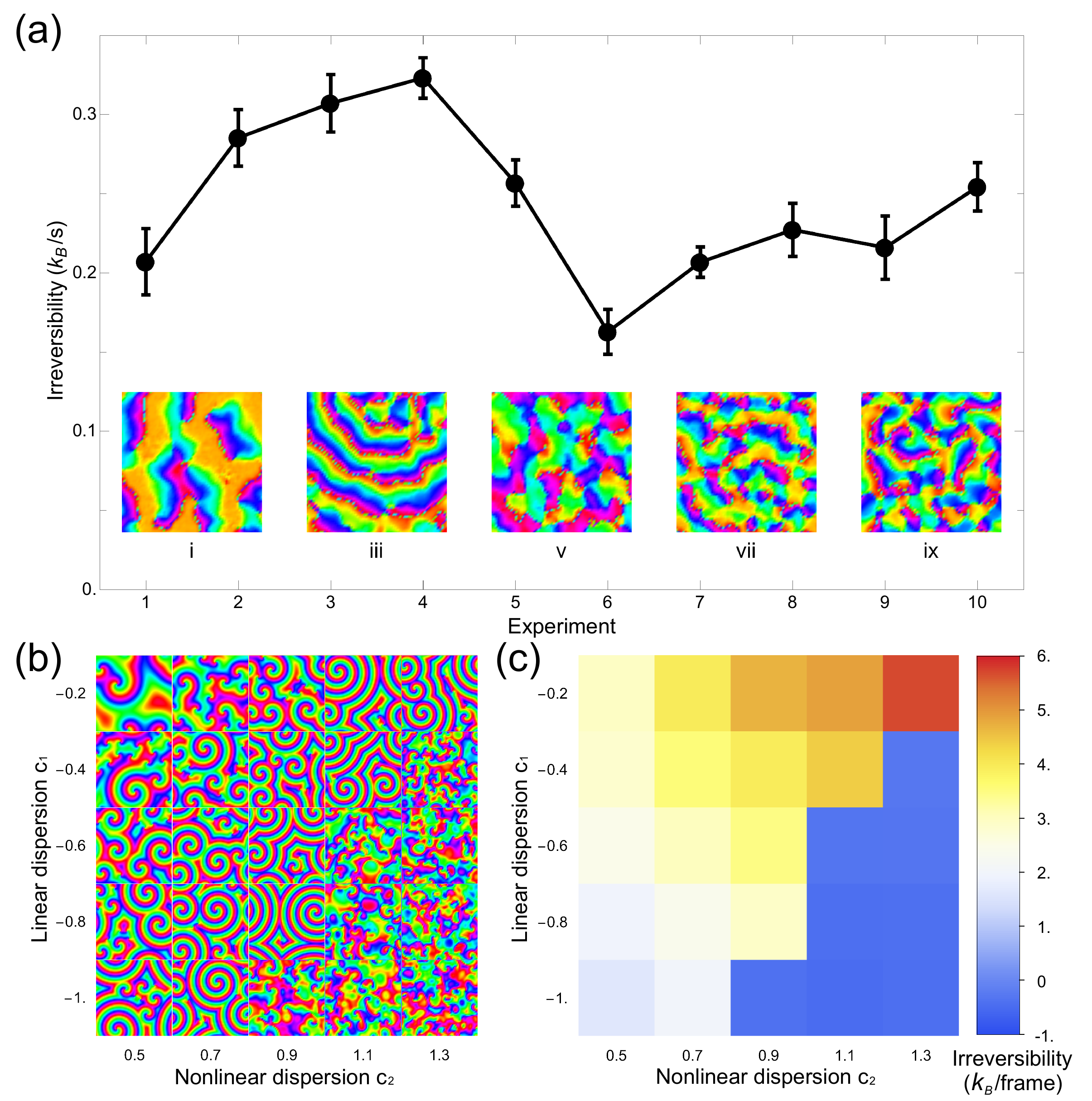}
  \caption{Irreversibility estimates distinguish dynamical regimes of Rho and CGL pattern formation. 
    (a) Irreversibility estimates increase with activity for experimental Rho states in the stable regime, but not upon transitioning to the chaotic regime. Error bars indicate standard errors of averages over training seeds (Methods and Fig.~\ref{fig:crop_comparison} in SM). Experimental Rho states are numbered by increasing effective kinetic energy (Methods and Fig.~\ref{fig:effective_kinetic_energies} in SM). Numbered insets show representative snapshots.
    (b) The CGL model exhibits stable and chaotic regimes. Simulation snapshots are shown for varying dispersions. The chaotic regime lacking spiral waves is shown in the lower right.
    (c) Stable-regime CGL simulations show irreversibility estimates increasing with dispersion parameters $c_1$ and $c_2$. Chaotic-regime CGL simulations show vanishing irreversibility estimates. Irreversibilities are estimated from FVAE latent trajectories for CGL simulations of varying dispersion parameters.
    }
  \label{fig:phase_digaram}
\end{figure*}

\section{Irreversibility estimates reveal dynamical phase transitions}
\begin{figure*}[h!]
    \centering
    \includegraphics[width=0.6\linewidth]{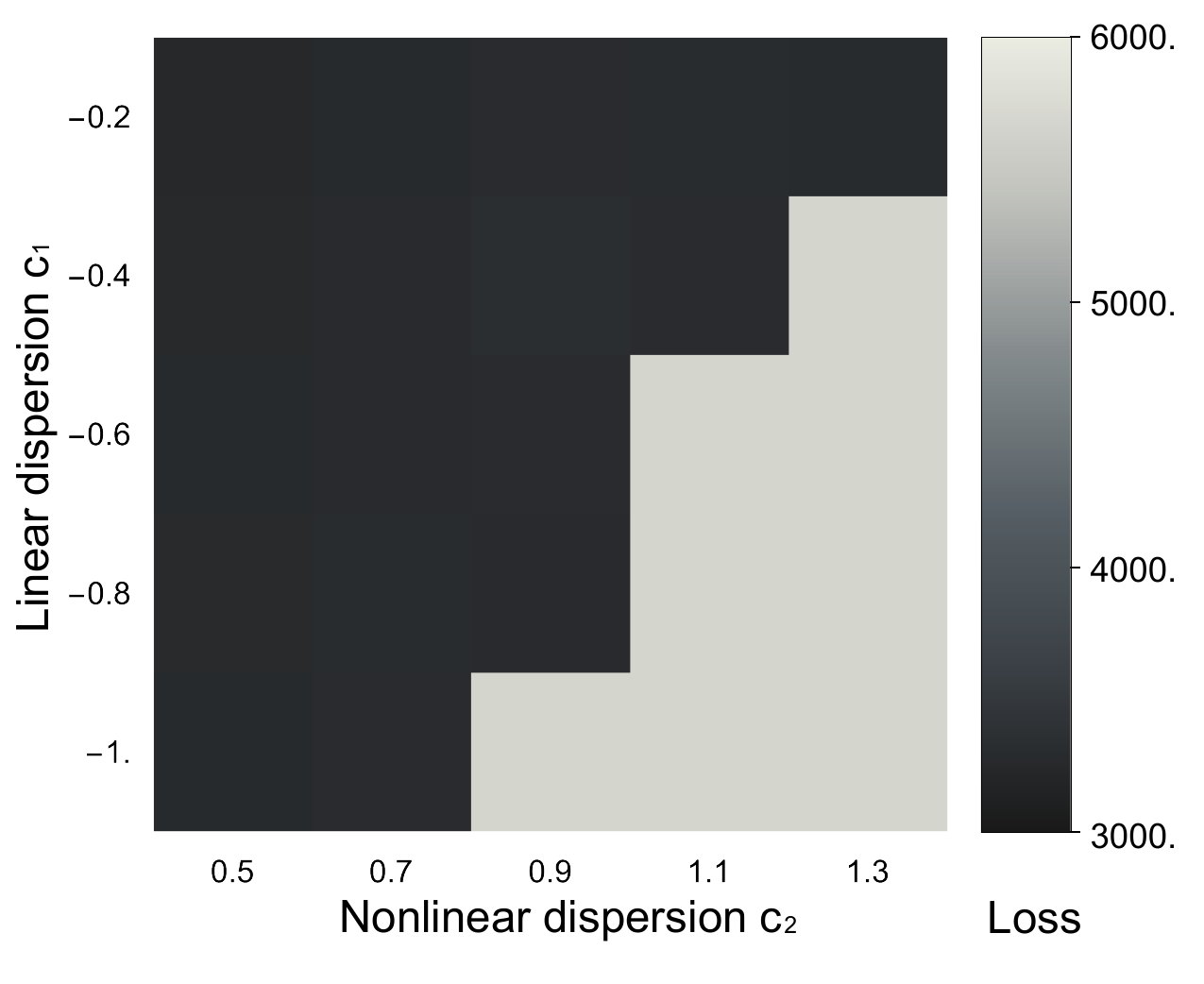}
    \caption{FVAE reconstruction losses vary little among models trained on stable CGL dynamics, but are much higher for models trained on turbulent CGL dynamics. Losses are averaged over models trained on three independent simulations. The trend is consistent with the observation that stable dynamics are better captured by FVAE than turbulent dynamics are.}
    \label{fig:loss_phase_diagram}
\end{figure*}

\begin{figure*}[h!]
    \centering
    \includegraphics[width=1\linewidth]{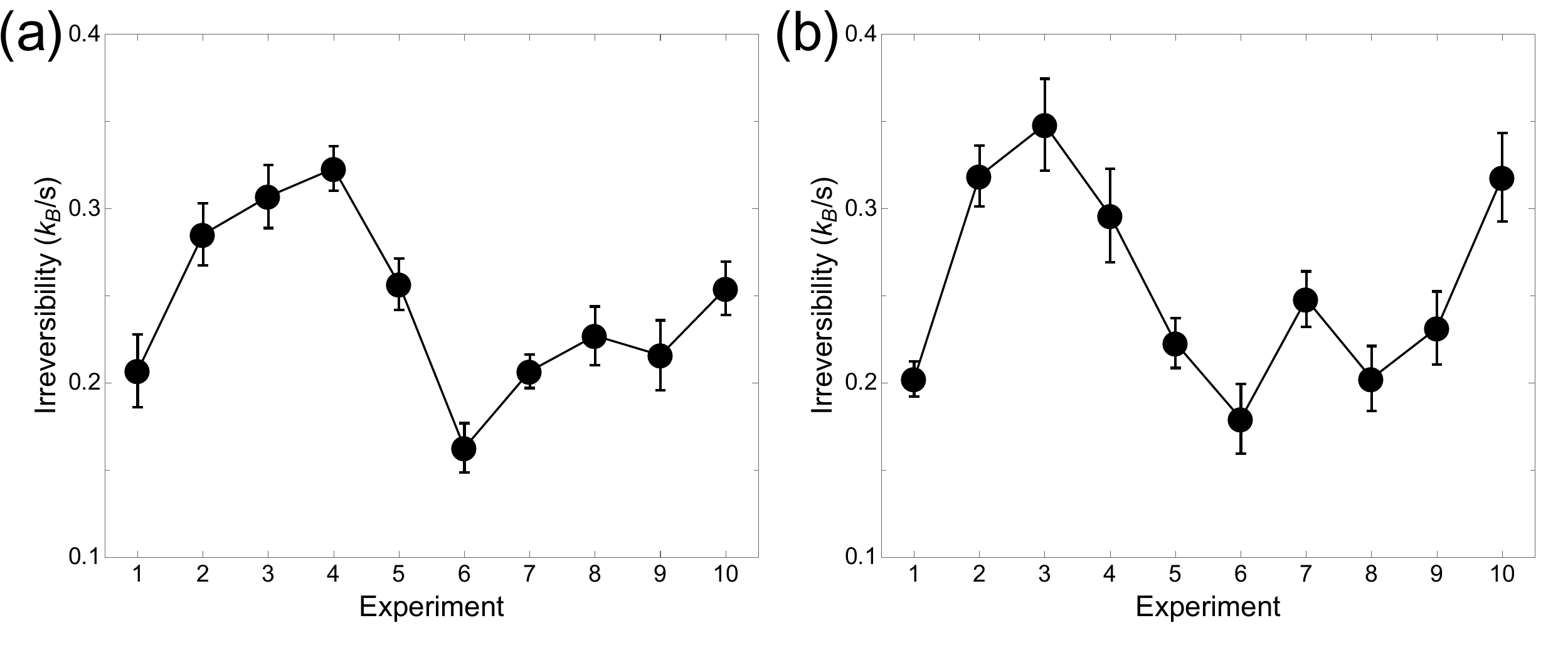}
    \caption{Irreversibility estimate trends are robust to choice of experimental crop. Irreversibility estimates are shown for non-overlapping crops of each experiment shown in Fig.~\ref{fig:phase_digaram}. Means and error bars denote averages and standard errors over training seeds.
    (a) Crops shown in Fig.~\ref{fig:phase_digaram}.
    (b) An additional crop of each experiment.}
    \label{fig:crop_comparison}
\end{figure*}

Further experiments on multiple ooctyes show irreversibility estimates initially increasing with pattern complexity across states numbered by effective kinetic energy, a measure of Rho activation rate (Methods and Fig.~\ref{fig:effective_kinetic_energies} in SM)~\cite{liu2021topological}.
However, irreversibility estimates decrease sharply above a critical Rho activation rate [curves in Figs.~\ref{fig:phase_digaram}(a) and \ref{fig:crop_comparison} in SM].
Rho patterns enter a chaotic regime, where stable spiral waves do not arise [insets in Fig.~\ref{fig:phase_digaram}(a)]~\cite{tan2020topological}.
This chaotic regime also occurs in CGL dynamics above a critical nonlinear dispersion $c_2$ or below a critical linear dispersion $c_1$ [Fig.~\ref{fig:phase_digaram}(b)].
The transition between stable spirals and chaotic turbulence can be detected through linear stability analysis of the CGL equation ~\cite{aranson2002world,chate1996phase}.
Interestingly, we recover this transition in estimated irreversibilities of patterns simulated at different dispersions [Fig.~\ref{fig:phase_digaram}(c)].
The nearly vanishing irreversibility estimates in the chaotic regime are counter-intuitive, but may be explained as follows:
1) In the stable regime, long-lived waves are a major and readily detected source of irreversibility.
However, the chaotic regime lacks structured wave-like motion: irreversibility might instead arise from higher-order correlations and non-exponential waiting times~\cite{lynn2022decomposing, skinner2021estimating}. 
Nonzero irreversibility estimates are thus difficult to obtain numerically.
2) The CGL equation describes a reaction-diffusion system with a reaction network not captured fully by the Rho phase field.
Chaos occurs at high nonlinear dispersion, which may correspond to more irreversibility arising in unobserved parts of the reaction pathway~\cite{falasco2018information,yu2021inverse}.
3) Chaotic patterns are unpredictable and susceptible to initial conditions not captured in a few latent dimensions. 
Reconstruction loss is greater for chaotic patterns than for stable patterns (Fig.~\ref{fig:loss_phase_diagram}).
Following these observations, we propose our irreversibility estimates as a dynamical order parameter distinguishing stable and chaotic regimes in nonlinear systems.
Lastly, at fixed nonlinear dispersion $c_2$, decreasing linear dispersion $c_1$ increases pattern complexity while decreasing irreversibility estimates [leftmost column, Fig.~\ref{fig:phase_digaram}(c)].
This behavior arises because irreversibility depends on both spatial and temporal structure: decreasing linear dispersion $c_1$ increases spatial frequency, but also decreases temporal frequency.

\section{Discussion}
In conclusion, we combine variational autoencoder networks with thermodynamic inference to robustly estimate and compare irreversibilities in spatiotemporally evolving biological patterns.
Our framework does not rely on prior knowledge of system dynamics, and we expect that it is applicable to general high-dimensional biological time series.
Resulting irreversibility estimates are necessarily lower bounds, as the ZM estimator requires coarse-graining and converges to the true irreversibility only in the limit of infinitely long time series encoding all dynamical degrees of freedom~\cite{roldan2012entropy}.
Phase-field patterns are of finite duration, encode neither absolute system size nor all degrees of freedom in underlying dynamics, and have FVAE reconstructions with loss.
However, our framework still reveals key features of simulated CGL and oocyte-Rho dynamics, including stability transitions between cell patterns and relative cell-activity levels.

\begin{figure*}[h!]
    \centering
    \includegraphics[width=1\linewidth]{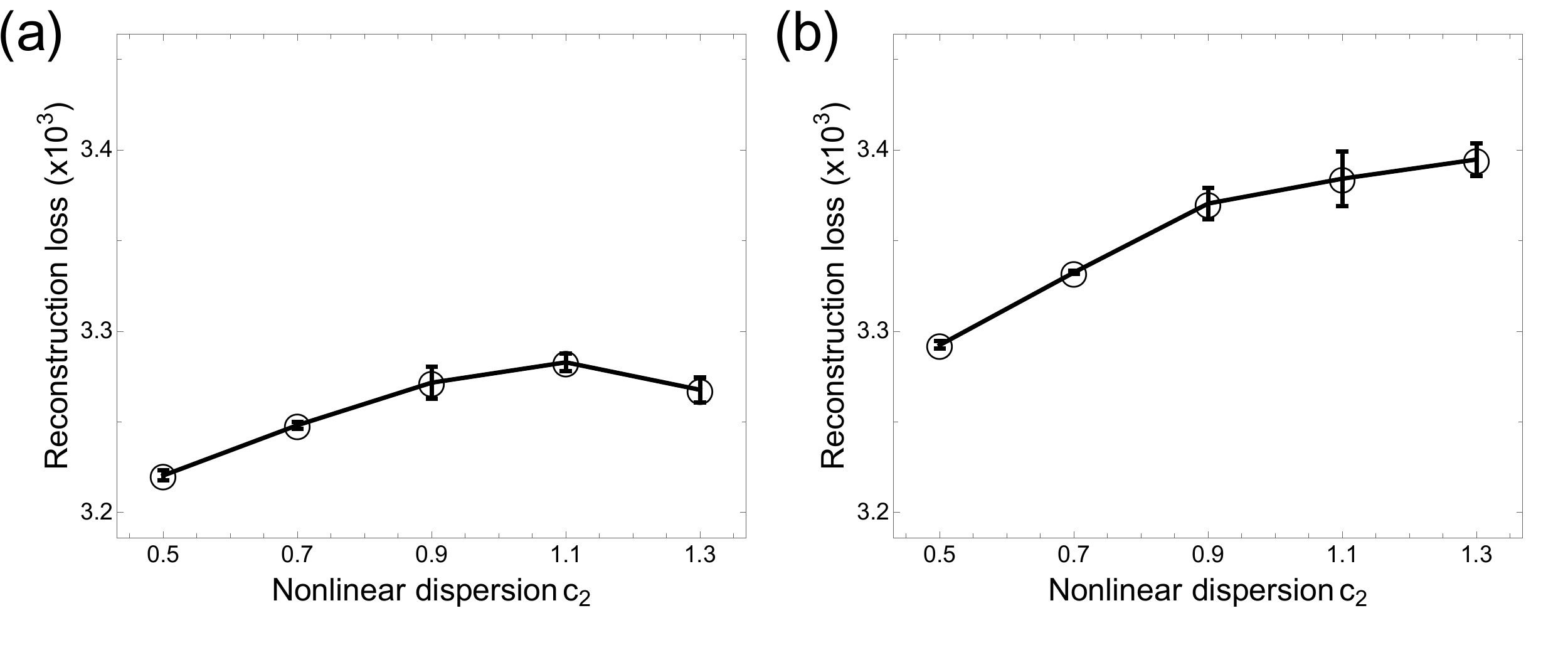}
    \caption{Reconstruction losses increase with pattern complexity in models trained on samples pooled from multiple patterns. Means and error bars denote averages and standard errors over three independent simulation datasets. (a) Reconstruction loss varies little with nonlinear dispersion $c_2$ at fixed linear dispersion $c_1=-0.2$ in models trained on all samples of one simulation. (b) Reconstruction loss increases monotonically with $c_2$ in models trained on $10^4$-sample datasets pooling the first 2000 samples from each of five simulations at $c_1=-0.2$.}
    \label{fig:pooled_bad}
\end{figure*}

With GPU acceleration, our framework is also highly computationally tractable.
Each model using a CGL simulation dataset trains for less than an hour, while each model using a Rho experiment dataset trains for less than five minutes.
A single model trained on samples pooled from multiple patterns has higher reconstruction loss on samples drawn from more complex patterns (Fig.~\ref{fig:pooled_bad} in SM).
To compare irreversibility estimates of patterns varying in complexity, we thus train separate models for each dataset and rescale latent trajectories before applying the ZM estimator.

With irreversibility ubiquitous out of equilibrium, our framework could rank activity levels while unveiling stability and potentially other dynamical properties in a broad range of living systems.
While limitations of Rho imaging require the use of complex-exponentiated phase fields, irreversibility could be estimated using models trained on raw intensity-field data obtained in high quality.
Additionally, our framework discovers efficient representations indexing complex dynamics by a few degrees of freedom, but physical interpretations of these latent dimensions are unknown.
FVAE neural-network weights capture large amounts of information about patterning systems not used in irreversibiility analyses.
Further physical interpretation of latent dimensions and FVAE weights might provide an intriguing avenue for understanding the origins of observed irreversibility.

\section{Methods}
\subsection{Rho data acquisition}
Experimental videos of Rho-GTPase patterns were obtained from previous studies~\cite{liu2021topological,tan2020topological}.
In brief, \textit{Patiria miniata} (bat sea star) oocytes were extracted and washed with filtered seawater.
Two constructs, eGFP-rGBD for labeling Rho-GTP molecules and Ect2-T808A-mCherry for generating excitable Rho-GTPase cortical patterns, were microinjected into the cytoplasm of the oocytes before incubation overnight at $15^\circ C$.
Microinjected oocytes were treated with 10 $\mu$M 1-methyl adenine solution to induce meiosis.
The oocyte in Fig.~\ref{fig:azide_results} was loaded into an open chamber constructed from glass and gas-permeable polymer (ibidi sticky-Slide) coverslips, while oocytes in Fig.~\ref{fig:phase_digaram} were loaded into customized polydimethylsiloxane (PDMS) chambers to minimize positional drift.
Time-lapse images of the ooctye in Fig.~\ref{fig:azide_results} were collected using $\times$60/NA 1.4 oil Plan Apochromat objective on a custom imaging setup.
Following half an hour of imaging with steady-state Rho patterning, the oocyte was treated with the metabolic inhibitor sodium azide NaN$_3$ (Sigma 71289).
Four consecutive half-hour patterns were then recorded.
Near-membrane Z-stack signals were collected for time-lapse confocal images of the oocyte in Fig.~\ref{fig:phase_digaram} using $\times$40/NA 1.3 oil Plan Apochromat objective with appropriate laser lines and emission filters.
The ten steady-state Rho patterns in Fig.~\ref{fig:phase_digaram} were recorded during ten contraction events over seven oocytes.

\subsection{Data processing}
We first obtain nonoverlapping $128 \times 128$-pixel crops from raw intensity data.
Phase field $\phi$ is calculated at each pixel over the entire 2D image. 
In order to minimize noise, we also performed background subtraction with a moving average over 15 frames.
We finally performed average pooling over $2 \times 2$-pixel kernels to generate $64 \times 64$-pixel phase-field frames $\bm{\phi}$.

\begin{figure*}[h!]
    \centering
    \includegraphics[width=1\linewidth]{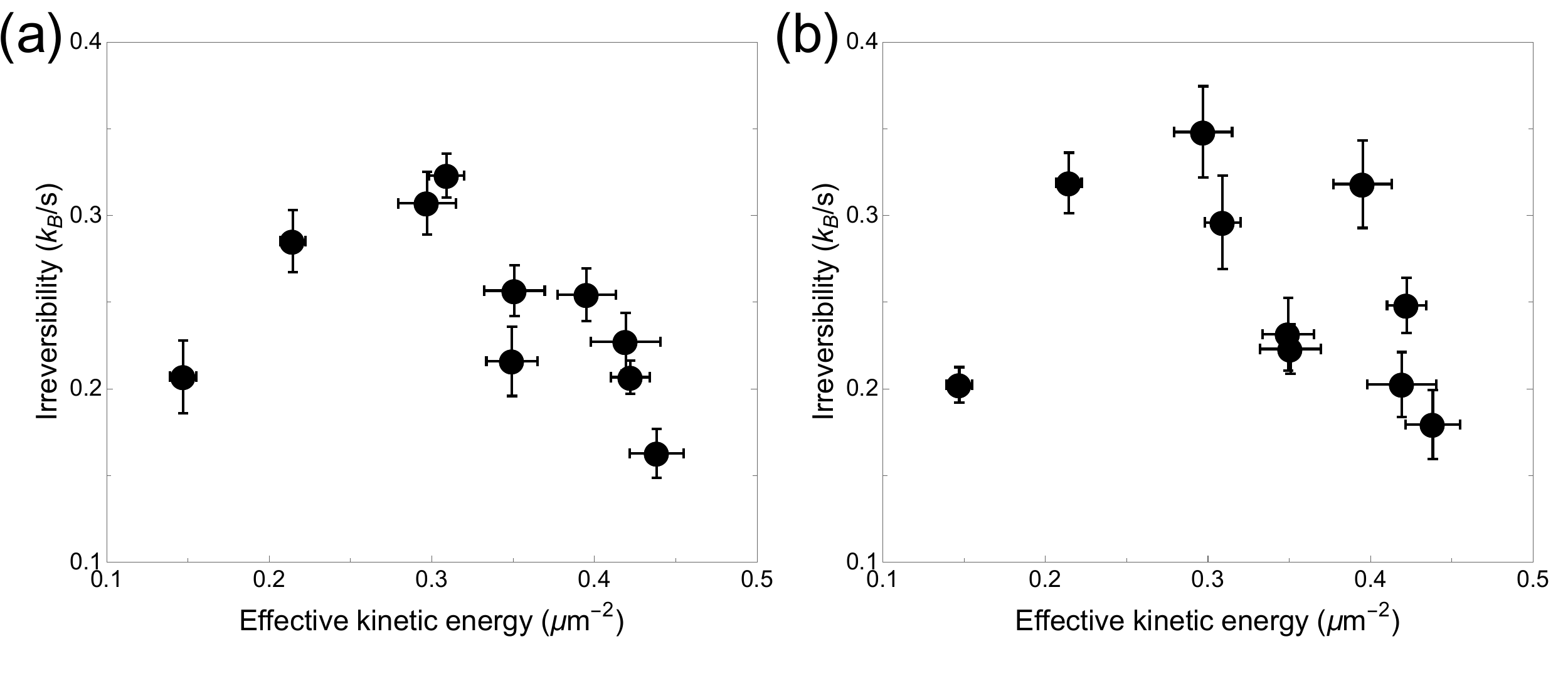}
    \caption{Irreversibility estimates reflect experimental cell-activity levels. Effective kinetic energy is a measure of cell-activity level. As in Fig.~\ref{fig:phase_digaram}(a), irreversibility estimates shown in each panel increase with cell-activity level in the stable regime (four points on the left) and are low in the turbulent regime (six points on the right). Irreversibility means and error bars denote averages and standard errors over training seeds; effective kinetic energy means and error bars denote averages and standard deviations over all pixels at all time points in the cell.
    (a) Crops shown in Fig.~\ref{fig:phase_digaram}.
    (b) An additional crop of each experiment.
    }
    \label{fig:effective_kinetic_energies}
\end{figure*}

To rank cell-activity levels, we calculated effective kinetic energies of different Rho phase patterns from corresponding phase-velocity fields: $\bm{V}_{\phi}=\nabla\phi$.
The effective kinetic energy is defined simply as $\langle |\bm{V}_{\phi}|^2\rangle$ with $\langle\cdot\rangle$ denoting an average over both space and time (Fig.~\ref{fig:effective_kinetic_energies} in SM).

\subsection{FVAE objective}
\label{loss}
The FVAE objective function (negative of the loss function Eq.~\ref{eq:objective}) is as previously described \cite{kim2018disentangling}.
In brief, we assume that $N$ observations $\bm{x}^{(i)} \in \bm{X}, i=1, \dots, N$ are generated by combining $K$ independent underlying factors of variation $\bm{f}=(f_1,\dots,f_K)$.
The FVAE uses real-valued latent vectors $\bm{z} \in \mathbb{R}^{d}$ to represent observations.
The generative model is defined by a standard Gaussian prior $p(\bm{z})=\mathcal{N}(0,I)$, where $I$ is a $d$-dimensional identity matrix.
For each observation, the encoder produces the mean $\mu_j(\bm{x})$ and variance $\sigma_j^2(\bm{x})$ of variational posterior $q_{\bm{\theta}}(\bm{z}|\bm{x})=\prod_{j=1}^{d}\mathcal{N}(z_j|\mu_j(\bm{x}),\sigma_j^2(\bm{x}))$ parameterized by neural network $\bm{\theta}$.
The decoder $p_{\bm{\psi}}(\bm{x}|\bm{z})$ is parameterized by neural network $\bm{\psi}$.
Considering all observations in the dataset, the distribution of latent representations is
\begin{equation}
q(\bm{z})=\mathbb{E}_{p_{\rm obs}(\bm{x})}[q(\bm{z}|\bm{x})]=\frac{1}{N}\sum_{i=1}^{N}q(\bm{z}|\bm{x}^{(i)}),
\end{equation}
where $p_{\rm obs}(\bm{x})$ is the empirical distribution.

In a standard VAE, the evidence lower bound objective (ELBO):
\begin{equation}
    \begin{aligned}
        \frac{1}{N}\sum_{i=1}^{N}\Bigl\{\mathbb{E}_{q(\bm{z}|\bm{x}^{(i)})}[\ln{p(\bm{x}^{(i)}|\bm{z})}]-\\KL[q(\bm{z}|\bm{x}^{(i)})||\mathcal{N}(0,I)]\Bigl\}
    \end{aligned}
\label{eq:ELBO}
\end{equation}
bounds the log-likelihood from below.
The first term of Eq.~\ref{eq:ELBO} is a negative reconstruction (binary cross entropy) loss, while the second term containing the Kullback-Leibler (KL) divergence
\begin{equation}
KL[p||q]=\mathbb{E}_{p}[\ln{\frac{p}{q}}]
\end{equation}
is a regularizer for model complexity. 

The FVAE objective function modifies the VAE objective in Eq.~\ref{eq:ELBO} by subtracting a total correlation (TC)
\begin{equation}
KL[q(\bm{z})||\bar{q}(\bm{z})]
\end{equation}
where
\begin{equation}
\bar{q}(\bm{z})=\prod_{j=1}^{d}q(z_j)
\end{equation}
to learn latent factors encoding complementary subsets of the $K$ mutually independent $\bm{f}$ \cite{watanabe1960information}. 
The TC penalizes dependence between latent dimensions as the KL divergence between the aggregate posterior $q(\bm{z})$ and the product of aggregate-posterior marginals $\bar{q}(\bm{z})$. 
Samples of $q(\bm{z})$ are obtained by sampling a minibatch of $q(\bm{z}|\bm{x}^{(i)})$. 
Samples of the product of aggregate-posterior marginals are obtained by randomly permuting each latent variable across a sampled minibatch of $q(\bm{z}|\bm{x}^{(i)})$, approximating $\bar{q}(\bm{z})$ in large minibatches \cite{arcones1992bootstrap}.
A discriminator $D$ training to distinguish between samples of $q(\bm{z})$ and $\bar{q}(\bm{z})$ outputs an estimate $D(\bm{z})$ that each sample belongs to $q(\bm{z})$.
The TC is thereby approximated as
\begin{equation}
	KL[q(\bm{z})||\bar{q}(\bm{z})]\approx\mathbb{E}_{q(\bm{z})}[\ln\frac{D(\bm{z})}{1-D(\bm{z})}]
\end{equation}
in computing the VAE loss function during joint training with the discriminator $D$ \cite{nguyen2010estimating, sugiyama2012density}.

\subsection{FVAE architecture}
\begin{figure*}[h!]
    \centering
    \includegraphics[width=1\linewidth]{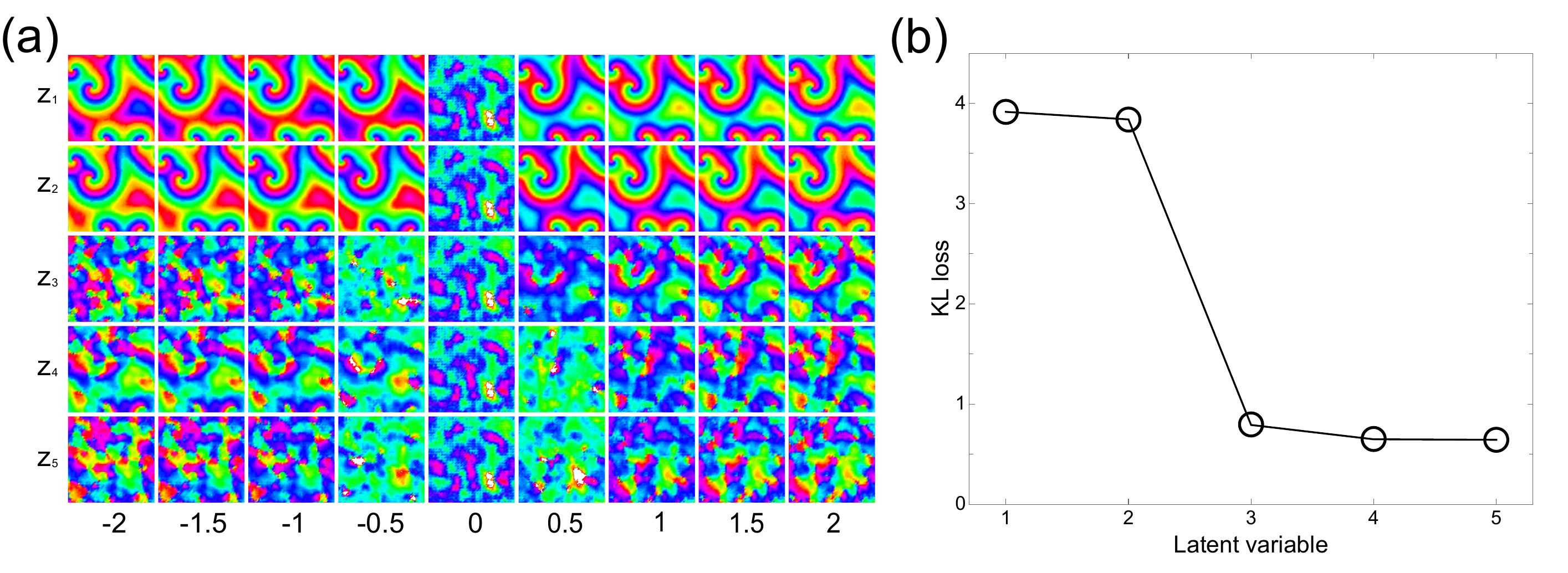}
    \caption{Models trained on simulations of stable CGL dynamics often converge on two-dimensional models.
    Latent dimensions are numbered by KL loss. 
    (a) Latent-dimension traversals change one variational-posterior mean while keeping others fixed at 0. Traversing latent dimensions $z_1$ and $z_2$ results in greater morphological change than traversing other latent dimensions. 
    (b) KL loss between the aggregate-posterior marginal and a normal prior is markedly greater for $z_1$ and $z_2$ than for other latent dimensions, indicating that these dimensions encode more information.}
    \label{fig:d_choice}
\end{figure*}

We adapted our architecture from open-source code~\cite{dubois2021disentangling, dupont2018learning}.
Each FVAE consists of a VAE and discriminator implemented in the PyTorch machine-learning package~\cite{paszke2019pytorch}.
The VAE has a feedforward architecture, with signals passing sequentially through the encoder, bottleneck, and decoder [Fig.~\ref{fig:schematic}(e)].
The encoder comprises four convolutional layers followed by two 256-unit linear layers.
The bottleneck comprises a 4-unit linear layer, encoding means and variances of variational posteriors in two-dimensional latent space, followed by a 2-unit linear layer, encoding latent-space vectors sampled from the variational posterior through reparameterization~\cite{kingma2013autoencoding}.
The decoder comprises two 128-unit linear layers followed by four transposed convolutional layers.
All convolutional and transposed convolutional layers have 1-pixel dilation, 1-pixel padding, 2-pixel stride, and $4\times4$-pixel kernel.
This architecture is largely as previously described \cite{burgess2018understanding}.
However, we use 4-unit and 2-unit bottleneck layers for two-dimensional latent spaces instead of 20-node and 10-node bottleneck layers for ten-dimensional latent spaces.
Models were initially trained with a larger number of latent dimensions (Fig.~\ref{fig:d_choice} in SM).
Models trained on simulations of stable CGL dynamics often converge on two-dimensional models: the number of latent dimensions was set to two to facilitate comparison between final models.
Moreover, we use a sigmoid activation in the final layer of the decoder.
The VAE otherwise uses ReLU activations.
The discriminator is a previously described perceptron with six 1000-unit layers using leaky ReLU activations of negative-domain slope 0.2, which outputs two logits as estimated unnormalized log-probabilities of inputs belonging to $q(\bm{z})$ and $\bar{q}(\bm{z})$ classes~\cite{kim2018disentangling}.

\subsection{FVAE training}
Phase-field frames $\bm{\phi}$ are $64\times 64$-pixel images of CGL or Rho phase-field dynamics sampled at evenly spaced points in time and saved in TIF format.
Inputs are $2\times 64\times 64$ tensors $\bm{x}$ transformed from $\bm{\phi}$ by applying
\begin{equation}
    x(\phi) = \frac{1}{2}e^{i\phi}+\frac{1}{2}+\frac{i}{2}
\label{eq:transform}
\end{equation}
to the phase field $\phi$ at each pixel, with the two channels of $\bm{x}$ containing real and imaginary parts of the $x$, respectively.
Each dataset consists of the inputs transformed from all frames of a single CGL or Rho phase-field video in addition to a $64\times 64$ vanishing-field reference frame.
Outputs are $2 \times 64 \times 64$ tensors, and are reconstructed inputs $\bm{x}$ or generated by decoding latent-space vectors.
Reconstructed phase-field frames $\bm{\phi}$ are obtained by inverting the transformation in Eq.~\ref{eq:transform} on complex-valued elements $x$ of reconstructed inputs $\bm{x}$.
During each training epoch, the dataset is randomly split into two equally sized minibatches, one for the VAE and the other for the discriminator.
FVAE loss (Eq. \ref{eq:objective}) is evaluated on the VAE minibatch, while mean binary cross-entropy loss is evaluated on the discriminator minibatch by normalizing output logits into class probabilities using the softmax function~\cite{goodfellow2016deep}.
Parameters are optimized using the Adam algorithm~\cite{kingma2015adam}; as previously described, moment-decay exponents are $\beta_1=0.9$ and $\beta_2=0.999$ for the VAE and $\beta_1=0.5$ and $\beta_2=0.9$ for the discriminator~\cite{kim2018disentangling}.
Each epoch, regression is performed on the FVAE losses over a hyperparameter “window” of previous epochs. 
Training stops once the resulting slope is not significantly less than zero at 95 percent confidence level for a hyperparameter “stop” number of consecutive epochs, suggesting that FVAE loss is no longer decreasing.

\begin{figure*}[h!]
    \centering
    \includegraphics[width=1\linewidth]{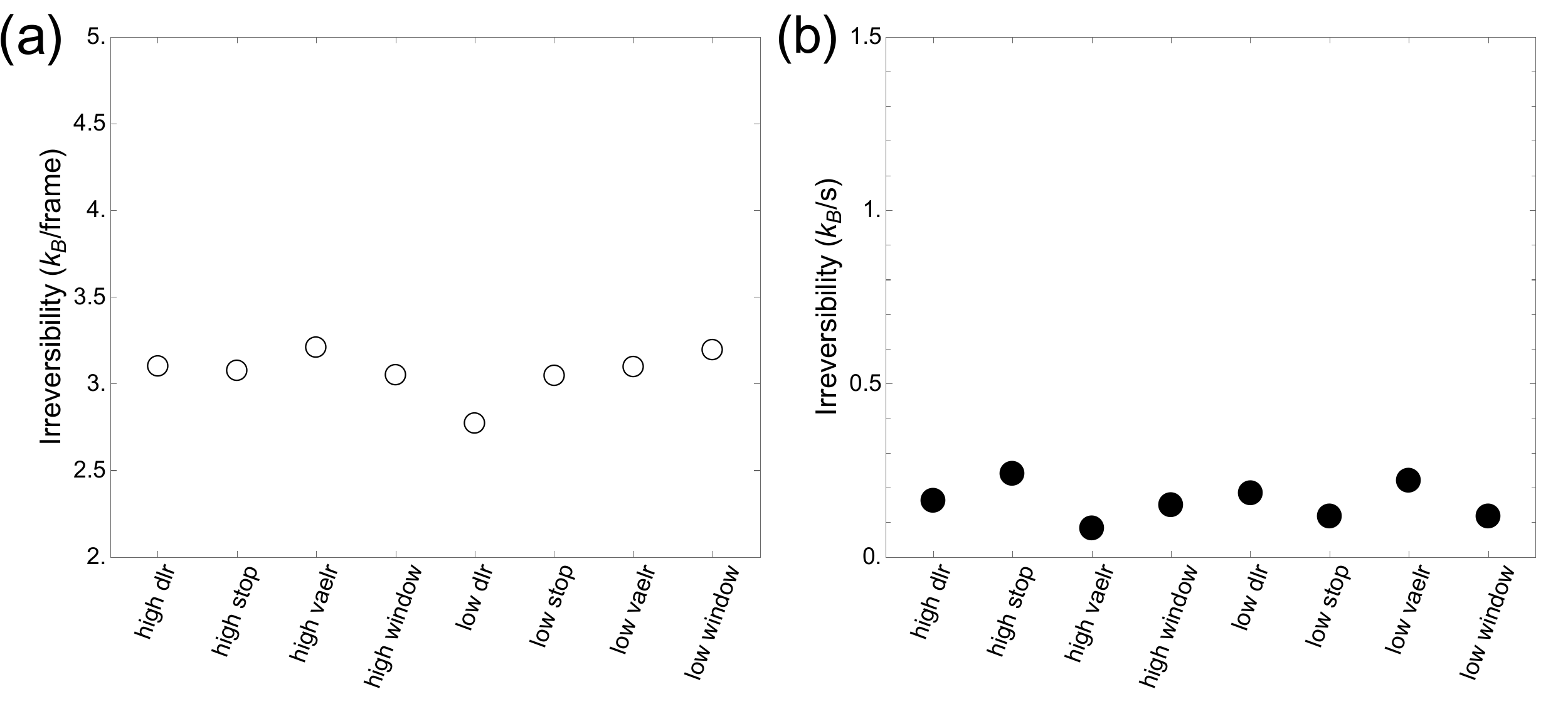}
    \caption{Irreversibility estimates vary little with choice of hyperparameters.
    Points labeled ``low'' for each hyperparameter have the hyperparameter halved from its default value (Table~\ref{tab:hyperparameters} in SM), while points labeled ``high'' for each hyperparameter have the hyperparameter doubled from its default value.
    (a) Irreversibilities of models trained on a simulation dataset with $c_1=-0.2$, $c_2=0.5$, timestep $0.1$, and initialization seed $0$.
    (b) Irreversibilities of models trained on an experiment dataset from crop $1$ of experiment $i$.}
    \label{fig:hyperparameters_vary}
\end{figure*}

Learning rates, window epochs, and stop epochs were selected by hyperparameter tuning~\cite{goodfellow2016deep}.
To determine whether tuned hyperparameters affect estimated irreversibilities, additional models were trained each with one tuned hyperparameter decreased or increased by a factor of two from its default value (Table~\ref{tab:hyperparameters} in SM).
Such models were trained for each tuned hyperparameter on a simulation dataset ($c_1=-0.2$, $c_2=0.5$, timestep $0.1$, and initialization seed $0$) and an experiment dataset (crop $1$ of experiment $i$): estimated irreversibilities vary little with choice of tuned hyperparameters (Fig.~\ref{fig:hyperparameters_vary} in SM).
Models of simulation datasets in main results used training seed 1234, with irreversibility estimates averaged over three independent simulations of each condition.
Models of experiment datasets in main results used training seeds 1234, 1243, 1324, 1342, 1423, 1432, 2134, 2143, 2314, 2341, 2413, and 2431, with irreversibility estimates averaged over the twelve models of each condition.
All models were trained on an Nvidia Titan RTX graphics card with CUDA driver.

\subsection{CGL phase-field simulations}
Complex Ginzburg-Landau phase-fields were simulated nondimensionally in MATLAB using ETD2 exponential time-differencing \cite{cox2002exponential, winterbottom2005complex}.
Multivariate Gaussian initial conditions of mean zero and covariance $10^{-4} I$, where $I$ denotes the identity matrix, were evolved on a $64\times 64$ grid with periodic boundary conditions for $10^6$ timesteps.
The last $10^4$ timesteps were retained.
For most simulations, linear and nonlinear dispersion parameters $c_1$ and $c_2$ were varied in increments of $0.2$ over $[-1.0, -0.2]$ and $[0.5, 1.3]$, respectively, at a timestep length of $0.10$.
In Fig.~\ref{fig:temporal_result}, simulations were sampled at timestep lengths $\{0.01, 0.05, 0.10, 0.50\}$ for the CGL equation with $c_1=-0.2$ and $c_2=0.5$.
The first three seeds to result in simulations reaching steady states (constant pixel oscillation envelopes) were used for each set of CGL dispersion parameters and timestep lengths.
In Fig.~\ref{fig:noise_result} and \ref{fig:spatial_result}(d), an additional Gaussian noise of mean zero and standard deviation $2\pi\xi$ was added independently to the phase field at each pixel of the timestep $0.10$ CGL simulations with $c_1=-0.2$, for $\xi$ ranging from $2^{-6}$ to $2^{-3}$ in powers of two.
Simulations were performed on a 3.3 GHz Quad-Core Intel Core i5 device.

\subsection{ZM compression estimator}
Our Ziv-Merhav compression estimator is as previously described~\cite{roldan2012entropy}.
For any time-series trajectory $\underline{\bm{z}} = \{\bm{z}^{(1)}, \bm{z}^{(2)},  \dots, \bm{z}^{(N)}\}$ and its reverse $\Tilde{\underline{\bm{z}}} = \{\bm{z}^{(N)}, \bm{z}^{(N-1)},  \dots, \bm{z}^{(1)}\},$ the ZM estimator is
\begin{equation}
    \hat{\dot{S}}_{\rm ZM} = \frac{1}{N}[c_{r}(\underline{\bm{z}}|\Tilde{\underline{\bm{z}}})\ln{N} - c(\underline{\bm{z}})\ln{c(\underline{\bm{z}})}].
    \label{eq:ZM}
\end{equation}
The first term in Eq.~\ref{eq:ZM} is the cross entropy rate, where $c_r(\underline{\bm{z}}|\Tilde{\underline{\bm{z}}})$ is defined as the length after parsing the forward trajectory $\underline{\bm{z}}$ by its reverse $\Tilde{\underline{\bm{z}}}$ using the Lempel-Ziv (LZ) algorithm~\cite{ziv1977universal}.
This term is also known as the cross-parsing length.
The second term is the Shannon entropy rate and $c(\underline{\bm{z}})$ denotes the length of $\underline{\bm{z}}$ after compressing with the LZ algorithm. 
To improve performance with limited data, we correct our estimator by applying it to half trajectories
\begin{equation}
    \begin{aligned}
        \hat{\dot{S}}_{\rm half} = \frac{2}{N}[c_{r}(\underline{\bm{z}}_{N/2}^{N}|\Tilde{\underline{\bm{z}}}_{1}^{N/2})\ln{N/2} -\\c(\underline{\bm{z}}_{N/2}^{N})\ln{c(\underline{\bm{z}}_{N/2}^{N})}]
    \end{aligned}
\end{equation}
and subtracting the asymptotically vanishing component 
\begin{equation}
    \hat{\dot{S}}_{\rm corr} = \hat{\dot{S}}_{\rm ZM} - \hat{\dot{S}}_{\rm half},
\end{equation}
again as previously described~\cite{roldan2012entropy}.

\begin{figure*}[h!]
    \centering
    \includegraphics[width=1\linewidth]{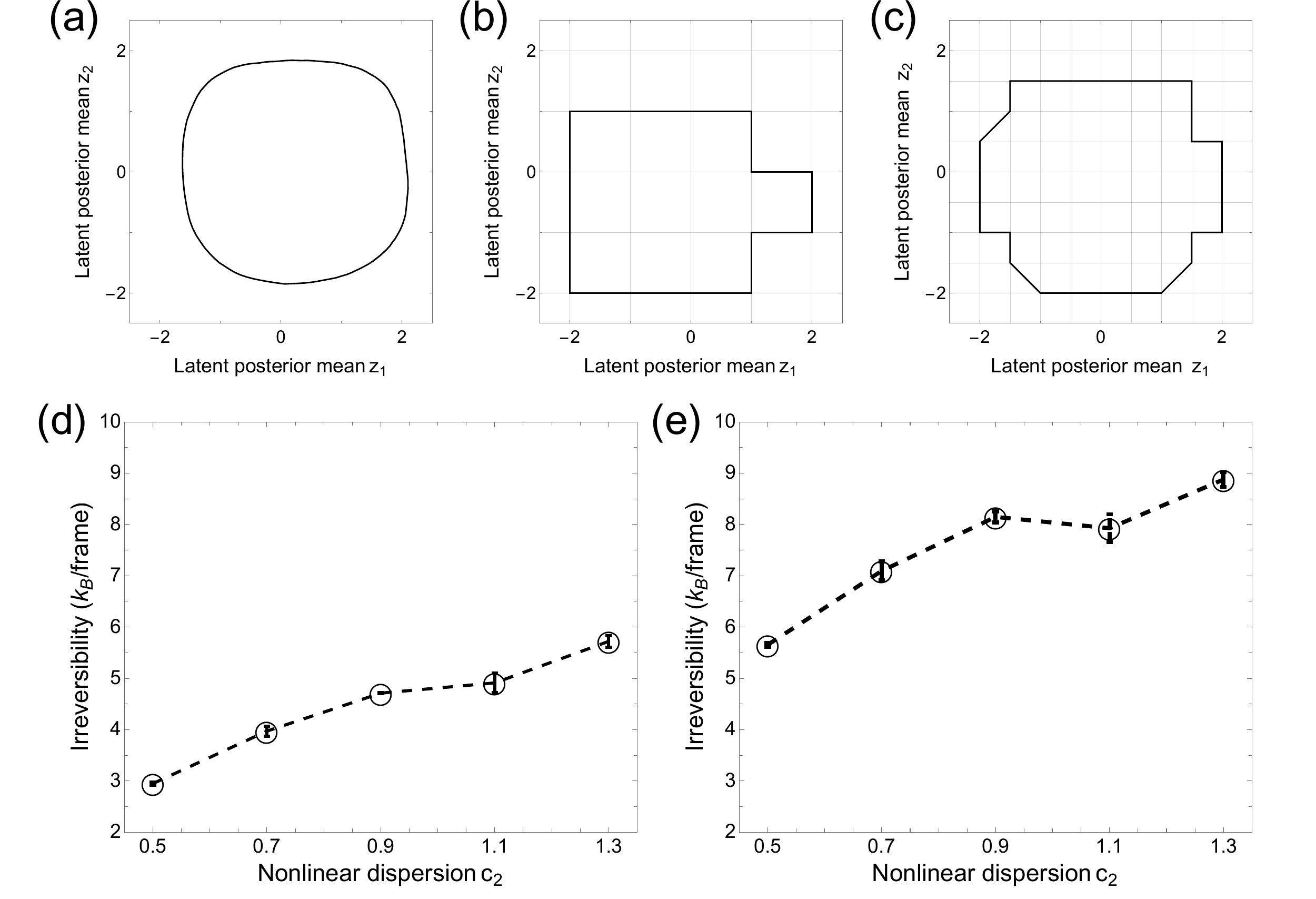}
    \caption{Finer discretizations yield higher irreversibility estimates. Irreversibility estimates increase with nonlinear dispersion $c_2$ and pattern complexity at fixed linear dispersion $c_1$.
    Irreversibility means and error bars denote averages and standard errors over three replicates.
    (a) The example latent trajectory from Fig.~\ref{fig:temporal_result}(a) (main text).
    (b) The trajectory in (a) discretized with $b=1$.
    (c) The trajectory in (a) discretized with $b=0.5$.
    (d) Estimated irreversibilities for the discretization shown in (b).
    (e) Estimated irreversibilities for the discretization shown in (c).
    }
    \label{fig:traj_coarse}
\end{figure*}

Since latent trajectories are of high precision, they are barely compressible.
In order to use the ZM compression estimator, we first discretize our latent trajectories using the floor function $\lfloor \bm{z}/b \rfloor$, with $b$ the parameter controlling discretization [Fig.~\ref{fig:traj_coarse}(a)-(c) in SM]. 
Although irreversibility estimates are higher for finer discretizations, the trend across regimes is preserved [Fig.~\ref{fig:traj_coarse}(d)-(e) in SM]. 
With enough data, different discretizations scale irreversibility estimates without altering their rankings. 
All main results are calculated with $b=1$.

\subsection{Latent-trajectory rescaling}
Since VAE exhibit a Lipschitz property (Eq.~\ref{eq:lipschitz}),
we rescale the latent trajectories of different patterns by the patterns' distances to a vanishing-field reference as
\begin{equation}
    \bm{z}_{\rm rescale} = \frac{\langle ||\bm{x} - \bm{x}_0 || \rangle}{5\langle ||\bm{z} - \bm{z}_0|| \rangle}\bm{z}.
\end{equation}
Here $\bm{x}_0$ is the reference and $\bm{z}_0$ is its position in a latent space after training.
$\langle\cdot\rangle$ denotes a time average over the entire simulation or experiment.

\section*{Data availability}
All data and code supporting this study are available for download at \url{https://doi.org/10.5281/zenodo.7734339} and \url{https://doi.org/10.5281/zenodo.7737963}, respectively.

\begin{acknowledgments}
We thank Jinghui Liu, Yu-Chen Chao, and Tzer Han Tan for help in data acquisition, Jordan M. Horowitz and Sarah E. Marzen for comments on the manuscript, and Hong-Hsing Liu for sharing compute resources.
This work was supported by National Science Foundation CAREER Grant No. PHYS-1848247 (to N.F.) and Alfred P. Sloan Foundation Grant G-2021-16758 (to N.F.).
\end{acknowledgments}

\section*{Author Contributions}
J.L., C.-W.J.L, and N.F. designed research. J.L., C.-W.J.L., and M.S. performed research. J.L., C.-W.J.L., and M.S. contributed new reagents/analytic tools. J.L. and C.-W.J.L. analyzed data. J.L., C.-W.J.L., and N.F. wrote the paper.

\section*{Appendix: Divergence of ZM Estimator on Deterministic Processes}
\begin{figure*}[htb!]
    \centering
    \includegraphics[width=1\linewidth]{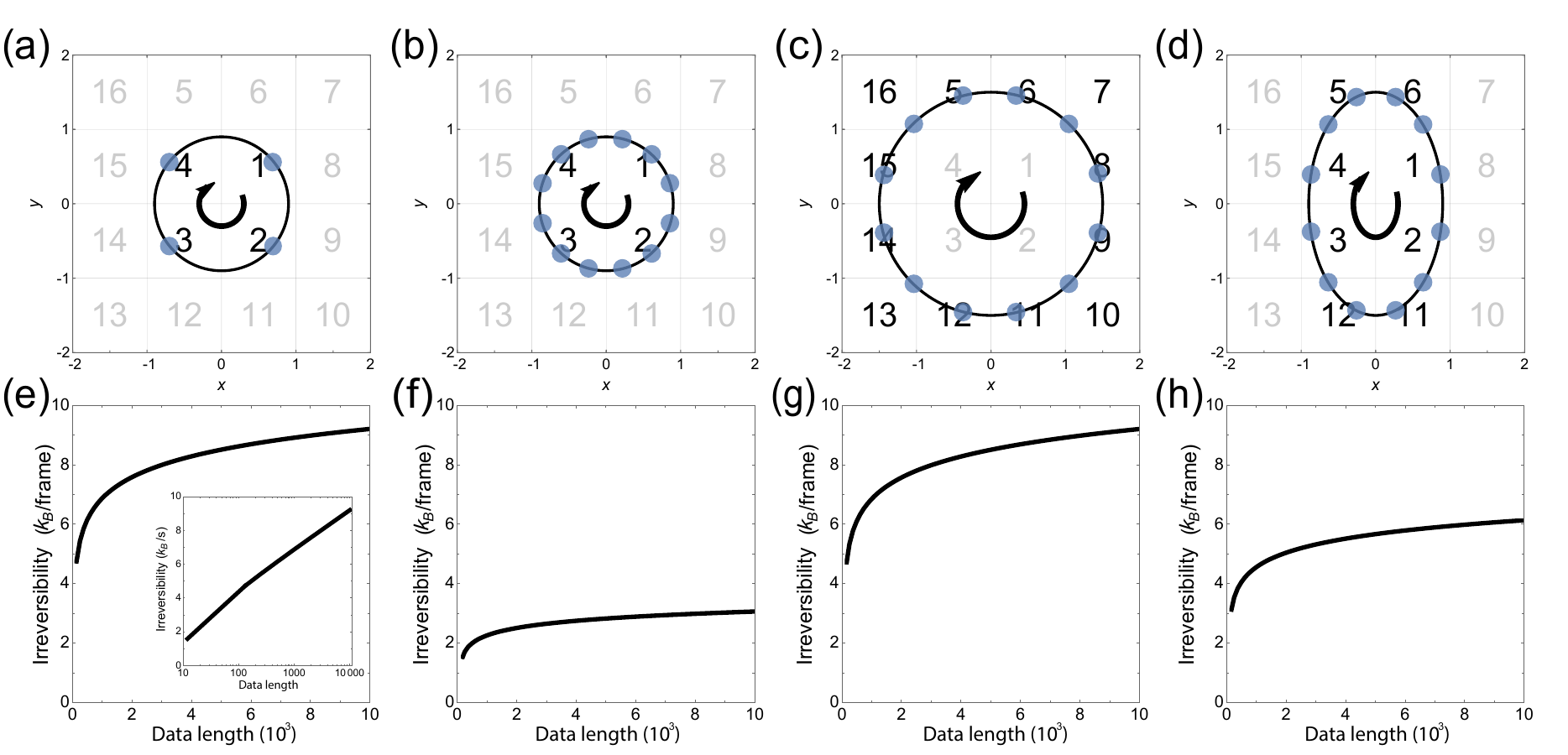}
    \caption{Irreversibility estimates can be compared between deterministic trajectories. 
    (a) Example cyclic trajectory. Arrow indicates direction and blue dots show points observed at a fixed sampling rate. Space is discretized into 16 bins of equal size. 
    (b) The same dynamics as in (a) but with speed decreased by a factor of three.
    (c) The same dynamics as in (b) but spatially dilated.
    (d) Example elliptical trajectory.
    (e) ZM irreversibility estimates calculated for the discretized series in (a) at different data lengths. Log-linear inset shows irreversibility increasing logarithmically with data length. 
    (f) ZM irreversibility estimates calculated for the discretized series in (b) at different data lengths.
    (g) ZM irreversibility estimates calculated for the discretized series in (c) at different data lengths.
    (h) ZM irreversibility estimates calculated for the discretized series in (d) at different data lengths.
    }
    \label{fig:diverge}
\end{figure*}

As CGL dynamics are deterministic, stochastic reverse processes are not observed. 
Irreversibilities of deterministic processes should diverge, raising the question of how such diverging irreversibilities can be compared.
Applied to a trajectory of finite length, the ZM estimator of irreversibility is finite~\cite{parrondo2009entropy,roldan2010estimating,roldan2012entropy}.
For deterministic trajectories, these estimates diverge logarithmically at different rates, allowing comparison.

Fig.~\ref{fig:diverge}(a) shows an example trajectory following deterministic dynamics:
\begin{equation}
    \begin{aligned}
        x &= r\cos{(\omega t - \pi/4)} \\
        y &= -r\sin{(\omega t - \pi/4)},
    \end{aligned}
    \label{eq:determin_example}
\end{equation}
with $r=0.9$, $\omega=\pi/2$, and time step 1.
As described in Methods, we discretize this trajectory with $b=1$ to apply the ZM estimator.
As a result, this trajectory can be labeled as a series of discrete indices $\underline{z}=\{1,2,3,4,1,2,3,4,1,2,3,4,\dots\}$.
For simplicity, we assume the series is finite with length $4N$ ($N \gg 4$) and its reverse is $\tilde{\underline{z}}=\{4,3,2,1,4,3,2,1,4,3,2,1,\dots\}$.
Plugging this series and its reverse into Eq.~\ref{eq:ZM}, we derive the cross parsing length $c_r(\underline{z}|\Tilde{\underline{z}})=4N$, the compression length $c(\underline{z}) \approx \log_2{4N}$, and the ZM compression estimator 
\begin{equation}
    \hat{\dot{S}}_{\rm ZM} \approx \ln{4N}- \frac{\log_2{4N}}{4N}.
\end{equation}
As $N \rightarrow \infty$, we can see that $\hat{\dot{S}}_{\rm ZM} \rightarrow \ln{4N}$ and diverges logarithmically as shown in Fig.~\ref{fig:diverge}(e).
However, if the dynamics in Eq.~\ref{eq:determin_example} are slower with $\omega = \pi/6$, the discretized trajectory becomes $\underline{z}=\{1,1,1,2,2,2,3,3,3,4,4,4,\dots\}$ [Fig.~\ref{fig:diverge}(b)].
The resulting ZM estimator $\hat{\dot{S}}_{\rm ZM} \rightarrow \frac{\ln{4N}}{3}$ [Fig.~\ref{fig:diverge}(f)].
Although the estimated irreversibility will eventually diverge, for a trajectory of the same finite length, it is one third that of the faster-evolving case shown in Fig.~\ref{fig:diverge}(a).
This property helps us to distinguish and compare different temporal structures in main text Fig.~\ref{fig:temporal_result}.

We can similarly compare different spatial structures. 
In Fig.~\ref{fig:diverge}(c), we dilated the trajectory in Fig.~\ref{fig:diverge}(b) with $\omega=\pi/6$ and $r=1.5$.
Indeed, the ZM estimate for the trajectory in Fig.~\ref{fig:diverge}(c) is greater than that for the trajectory in Fig.~\ref{fig:diverge}(b) with the same trajectory length and discretization [Fig.~\ref{fig:diverge}(g)].
More precisely, the long-time finite ZM estimate shown in Fig.~\ref{fig:diverge}(g) is three times that shown in Fig.~\ref{fig:diverge}(f).
This is because, though the trajectories in Fig.~\ref{fig:diverge}(b) and Fig.~\ref{fig:diverge}(c) have the same period, the latter trajectory visits three times as many distinct discretized positions as the former: it is one third as compressible using its reverse.
Moreover, the finite ZM estimator also captures different trajectory geometries. 
The elliptical dynamics shown in Fig.~\ref{fig:diverge}(d),  which follow $x = 0.9\cos{( \pi t/6 - \pi/4)}$ and $y = -1.5\sin{( \pi t/6 - \pi/4)}$, will evolve at different speeds along the trajectory. 
The spatially varying trajectory speed increases the compressibility of the trajectory, as only eight distinct discretized positions are visited, and is manifested in the ZM estimator [Fig.~\ref{fig:diverge}(h)]. 
As a result, we are able to compare the irreversibilities of different spatial structures in main text Fig.~\ref{fig:spatial_result}.

Altogether, we conclude that even for deterministic processes, the finite-trajectory ZM estimator still distinguishes and enables comparisons between dynamics.

\bibliography{main}

\newcommand{\noopsort}[1]{} \newcommand{\printfirst}[2]{#1} \newcommand{\singleletter}[1]{#1} \newcommand{\switchargs}[2]{#2#1}
\begin{thebibliography}{61}%
\makeatletter
\providecommand \@ifxundefined [1]{%
 \@ifx{#1\undefined}
}%
\providecommand \@ifnum [1]{%
 \ifnum #1\expandafter \@firstoftwo
 \else \expandafter \@secondoftwo
 \fi
}%
\providecommand \@ifx [1]{%
 \ifx #1\expandafter \@firstoftwo
 \else \expandafter \@secondoftwo
 \fi
}%
\providecommand \natexlab [1]{#1}%
\providecommand \enquote  [1]{``#1''}%
\providecommand \bibnamefont  [1]{#1}%
\providecommand \bibfnamefont [1]{#1}%
\providecommand \citenamefont [1]{#1}%
\providecommand \href@noop [0]{\@secondoftwo}%
\providecommand \href [0]{\begingroup \@sanitize@url \@href}%
\providecommand \@href[1]{\@@startlink{#1}\@@href}%
\providecommand \@@href[1]{\endgroup#1\@@endlink}%
\providecommand \@sanitize@url [0]{\catcode `\\12\catcode `\$12\catcode `\&12\catcode `\#12\catcode `\^12\catcode `\_12\catcode `\%12\relax}%
\providecommand \@@startlink[1]{}%
\providecommand \@@endlink[0]{}%
\providecommand \url  [0]{\begingroup\@sanitize@url \@url }%
\providecommand \@url [1]{\endgroup\@href {#1}{\urlprefix }}%
\providecommand \urlprefix  [0]{URL }%
\providecommand \Eprint [0]{\href }%
\providecommand \doibase [0]{https://doi.org/}%
\providecommand \selectlanguage [0]{\@gobble}%
\providecommand \bibinfo  [0]{\@secondoftwo}%
\providecommand \bibfield  [0]{\@secondoftwo}%
\providecommand \translation [1]{[#1]}%
\providecommand \BibitemOpen [0]{}%
\providecommand \bibitemStop [0]{}%
\providecommand \bibitemNoStop [0]{.\EOS\space}%
\providecommand \EOS [0]{\spacefactor3000\relax}%
\providecommand \BibitemShut  [1]{\csname bibitem#1\endcsname}%
\let\auto@bib@innerbib\@empty
\bibitem [{\citenamefont {Needleman}\ and\ \citenamefont {Dogic}(2017)}]{needleman2017active}%
  \BibitemOpen
  \bibfield  {author} {\bibinfo {author} {\bibfnamefont {D.}~\bibnamefont {Needleman}}\ and\ \bibinfo {author} {\bibfnamefont {Z.}~\bibnamefont {Dogic}},\ }\href {https://doi.org/10.1038/natrevmats.2017.48} {\bibfield  {journal} {\bibinfo  {journal} {Nature Reviews Materials}\ }\textbf {\bibinfo {volume} {2}},\ \bibinfo {pages} {17048} (\bibinfo {year} {2017})}\BibitemShut {NoStop}%
\bibitem [{\citenamefont {Gnesotto}\ \emph {et~al.}(2018)\citenamefont {Gnesotto}, \citenamefont {Mura}, \citenamefont {Gladrow},\ and\ \citenamefont {Broedersz}}]{gnesotto2018broken}%
  \BibitemOpen
  \bibfield  {author} {\bibinfo {author} {\bibfnamefont {F.}~\bibnamefont {Gnesotto}}, \bibinfo {author} {\bibfnamefont {F.}~\bibnamefont {Mura}}, \bibinfo {author} {\bibfnamefont {J.}~\bibnamefont {Gladrow}},\ and\ \bibinfo {author} {\bibfnamefont {C.}~\bibnamefont {Broedersz}},\ }\href {https://doi.org/10.1088/1361-6633/aab3ed} {\bibfield  {journal} {\bibinfo  {journal} {Reports on Progress in Physics}\ }\textbf {\bibinfo {volume} {81}},\ \bibinfo {pages} {066601} (\bibinfo {year} {2018})}\BibitemShut {NoStop}%
\bibitem [{\citenamefont {Marchetti}\ \emph {et~al.}(2013)\citenamefont {Marchetti}, \citenamefont {Joanny}, \citenamefont {Ramaswamy}, \citenamefont {Liverpool}, \citenamefont {Prost}, \citenamefont {Rao},\ and\ \citenamefont {Simha}}]{marchetti2013hydrodynamics}%
  \BibitemOpen
  \bibfield  {author} {\bibinfo {author} {\bibfnamefont {M.~C.}\ \bibnamefont {Marchetti}}, \bibinfo {author} {\bibfnamefont {J.-F.}\ \bibnamefont {Joanny}}, \bibinfo {author} {\bibfnamefont {S.}~\bibnamefont {Ramaswamy}}, \bibinfo {author} {\bibfnamefont {T.~B.}\ \bibnamefont {Liverpool}}, \bibinfo {author} {\bibfnamefont {J.}~\bibnamefont {Prost}}, \bibinfo {author} {\bibfnamefont {M.}~\bibnamefont {Rao}},\ and\ \bibinfo {author} {\bibfnamefont {R.~A.}\ \bibnamefont {Simha}},\ }\href {https://doi.org/10.1103/RevModPhys.85.1143} {\bibfield  {journal} {\bibinfo  {journal} {Reviews of Modern Physics}\ }\textbf {\bibinfo {volume} {85}},\ \bibinfo {pages} {1143} (\bibinfo {year} {2013})}\BibitemShut {NoStop}%
\bibitem [{\citenamefont {Seifert}(2012)}]{seifert2012stochastic}%
  \BibitemOpen
  \bibfield  {author} {\bibinfo {author} {\bibfnamefont {U.}~\bibnamefont {Seifert}},\ }\href {https://doi.org/10.1088/0034-4885/75/12/126001} {\bibfield  {journal} {\bibinfo  {journal} {Reports on Progress in Physics}\ }\textbf {\bibinfo {volume} {75}},\ \bibinfo {pages} {126001} (\bibinfo {year} {2012})}\BibitemShut {NoStop}%
\bibitem [{\citenamefont {Murugan}\ and\ \citenamefont {Vaikuntanathan}(2016)}]{murugan2016biological}%
  \BibitemOpen
  \bibfield  {author} {\bibinfo {author} {\bibfnamefont {A.}~\bibnamefont {Murugan}}\ and\ \bibinfo {author} {\bibfnamefont {S.}~\bibnamefont {Vaikuntanathan}},\ }\href {https://doi.org/10.1007/s10955-015-1445-0} {\bibfield  {journal} {\bibinfo  {journal} {Journal of Statistical Physics}\ }\textbf {\bibinfo {volume} {162}},\ \bibinfo {pages} {1183} (\bibinfo {year} {2016})}\BibitemShut {NoStop}%
\bibitem [{\citenamefont {Crooks}(1999)}]{crooks1999entropy}%
  \BibitemOpen
  \bibfield  {author} {\bibinfo {author} {\bibfnamefont {G.~E.}\ \bibnamefont {Crooks}},\ }\href {https://doi.org/10.1103/PhysRevE.60.2721} {\bibfield  {journal} {\bibinfo  {journal} {Physical Review E}\ }\textbf {\bibinfo {volume} {60}},\ \bibinfo {pages} {2721} (\bibinfo {year} {1999})}\BibitemShut {NoStop}%
\bibitem [{\citenamefont {Luposchainsky}\ and\ \citenamefont {Hinrichsen}(2013)}]{luposchainsky2013entropy}%
  \BibitemOpen
  \bibfield  {author} {\bibinfo {author} {\bibfnamefont {D.}~\bibnamefont {Luposchainsky}}\ and\ \bibinfo {author} {\bibfnamefont {H.}~\bibnamefont {Hinrichsen}},\ }\href {https://doi.org/10.1007/s10955-013-0863-0} {\bibfield  {journal} {\bibinfo  {journal} {Journal of Statistical Physics}\ }\textbf {\bibinfo {volume} {153}},\ \bibinfo {pages} {828} (\bibinfo {year} {2013})}\BibitemShut {NoStop}%
\bibitem [{\citenamefont {Mabillard}\ \emph {et~al.}(2023)\citenamefont {Mabillard}, \citenamefont {Weber},\ and\ \citenamefont {J\"ulicher}}]{mabillard2023heat}%
  \BibitemOpen
  \bibfield  {author} {\bibinfo {author} {\bibfnamefont {J.}~\bibnamefont {Mabillard}}, \bibinfo {author} {\bibfnamefont {C.~A.}\ \bibnamefont {Weber}},\ and\ \bibinfo {author} {\bibfnamefont {F.}~\bibnamefont {J\"ulicher}},\ }\href {https://doi.org/10.1103/PhysRevE.107.014118} {\bibfield  {journal} {\bibinfo  {journal} {Physical Review E}\ }\textbf {\bibinfo {volume} {107}},\ \bibinfo {pages} {014118} (\bibinfo {year} {2023})}\BibitemShut {NoStop}%
\bibitem [{\citenamefont {Battle}\ \emph {et~al.}(2016)\citenamefont {Battle}, \citenamefont {Broedersz}, \citenamefont {Fakhri}, \citenamefont {Geyer}, \citenamefont {Howard}, \citenamefont {Schmidt},\ and\ \citenamefont {MacKintosh}}]{battle2016broken}%
  \BibitemOpen
  \bibfield  {author} {\bibinfo {author} {\bibfnamefont {C.}~\bibnamefont {Battle}}, \bibinfo {author} {\bibfnamefont {C.~P.}\ \bibnamefont {Broedersz}}, \bibinfo {author} {\bibfnamefont {N.}~\bibnamefont {Fakhri}}, \bibinfo {author} {\bibfnamefont {V.~F.}\ \bibnamefont {Geyer}}, \bibinfo {author} {\bibfnamefont {J.}~\bibnamefont {Howard}}, \bibinfo {author} {\bibfnamefont {C.~F.}\ \bibnamefont {Schmidt}},\ and\ \bibinfo {author} {\bibfnamefont {F.~C.}\ \bibnamefont {MacKintosh}},\ }\href {https://doi.org/10.1126/science.aac8167} {\bibfield  {journal} {\bibinfo  {journal} {Science}\ }\textbf {\bibinfo {volume} {352}},\ \bibinfo {pages} {604} (\bibinfo {year} {2016})}\BibitemShut {NoStop}%
\bibitem [{\citenamefont {Seif}\ \emph {et~al.}(2021)\citenamefont {Seif}, \citenamefont {Hafezi},\ and\ \citenamefont {Jarzynski}}]{seif2021machine}%
  \BibitemOpen
  \bibfield  {author} {\bibinfo {author} {\bibfnamefont {A.}~\bibnamefont {Seif}}, \bibinfo {author} {\bibfnamefont {M.}~\bibnamefont {Hafezi}},\ and\ \bibinfo {author} {\bibfnamefont {C.}~\bibnamefont {Jarzynski}},\ }\href {https://doi.org/10.1038/s41567-020-1018-2} {\bibfield  {journal} {\bibinfo  {journal} {Nature Physics}\ }\textbf {\bibinfo {volume} {17}},\ \bibinfo {pages} {105} (\bibinfo {year} {2021})}\BibitemShut {NoStop}%
\bibitem [{\citenamefont {Parrondo}\ \emph {et~al.}(2009)\citenamefont {Parrondo}, \citenamefont {Van~den Broeck},\ and\ \citenamefont {Kawai}}]{parrondo2009entropy}%
  \BibitemOpen
  \bibfield  {author} {\bibinfo {author} {\bibfnamefont {J.~M.~R.}\ \bibnamefont {Parrondo}}, \bibinfo {author} {\bibfnamefont {C.}~\bibnamefont {Van~den Broeck}},\ and\ \bibinfo {author} {\bibfnamefont {R.}~\bibnamefont {Kawai}},\ }\href {https://doi.org/10.1088/1367-2630/11/7/073008} {\bibfield  {journal} {\bibinfo  {journal} {New Journal of Physics}\ }\textbf {\bibinfo {volume} {11}},\ \bibinfo {pages} {073008} (\bibinfo {year} {2009})}\BibitemShut {NoStop}%
\bibitem [{\citenamefont {Li}\ \emph {et~al.}(2019)\citenamefont {Li}, \citenamefont {Horowitz}, \citenamefont {Gingrich},\ and\ \citenamefont {Fakhri}}]{li2019quantifying}%
  \BibitemOpen
  \bibfield  {author} {\bibinfo {author} {\bibfnamefont {J.}~\bibnamefont {Li}}, \bibinfo {author} {\bibfnamefont {J.~M.}\ \bibnamefont {Horowitz}}, \bibinfo {author} {\bibfnamefont {T.~R.}\ \bibnamefont {Gingrich}},\ and\ \bibinfo {author} {\bibfnamefont {N.}~\bibnamefont {Fakhri}},\ }\href {https://doi.org/10.1038/s41467-019-09631-x} {\bibfield  {journal} {\bibinfo  {journal} {Nature Communications}\ }\textbf {\bibinfo {volume} {10}},\ \bibinfo {pages} {1} (\bibinfo {year} {2019})}\BibitemShut {NoStop}%
\bibitem [{\citenamefont {Tan}\ \emph {et~al.}(2022)\citenamefont {Tan}, \citenamefont {Mietke}, \citenamefont {Li}, \citenamefont {Chen}, \citenamefont {Higinbotham}, \citenamefont {Foster}, \citenamefont {Gokhale}, \citenamefont {Dunkel},\ and\ \citenamefont {Fakhri}}]{tan2022odd}%
  \BibitemOpen
  \bibfield  {author} {\bibinfo {author} {\bibfnamefont {T.~H.}\ \bibnamefont {Tan}}, \bibinfo {author} {\bibfnamefont {A.}~\bibnamefont {Mietke}}, \bibinfo {author} {\bibfnamefont {J.}~\bibnamefont {Li}}, \bibinfo {author} {\bibfnamefont {Y.}~\bibnamefont {Chen}}, \bibinfo {author} {\bibfnamefont {H.}~\bibnamefont {Higinbotham}}, \bibinfo {author} {\bibfnamefont {P.~J.}\ \bibnamefont {Foster}}, \bibinfo {author} {\bibfnamefont {S.}~\bibnamefont {Gokhale}}, \bibinfo {author} {\bibfnamefont {J.}~\bibnamefont {Dunkel}},\ and\ \bibinfo {author} {\bibfnamefont {N.}~\bibnamefont {Fakhri}},\ }\href {https://doi.org/10.1038/s41586-022-04889-6} {\bibfield  {journal} {\bibinfo  {journal} {Nature}\ }\textbf {\bibinfo {volume} {607}},\ \bibinfo {pages} {287} (\bibinfo {year} {2022})}\BibitemShut {NoStop}%
\bibitem [{\citenamefont {Tan}\ \emph {et~al.}(2021)\citenamefont {Tan}, \citenamefont {Watson}, \citenamefont {Chao}, \citenamefont {Li}, \citenamefont {Gingrich}, \citenamefont {Horowitz},\ and\ \citenamefont {Fakhri}}]{tan2021scale}%
  \BibitemOpen
  \bibfield  {author} {\bibinfo {author} {\bibfnamefont {T.~H.}\ \bibnamefont {Tan}}, \bibinfo {author} {\bibfnamefont {G.~A.}\ \bibnamefont {Watson}}, \bibinfo {author} {\bibfnamefont {Y.-C.}\ \bibnamefont {Chao}}, \bibinfo {author} {\bibfnamefont {J.}~\bibnamefont {Li}}, \bibinfo {author} {\bibfnamefont {T.~R.}\ \bibnamefont {Gingrich}}, \bibinfo {author} {\bibfnamefont {J.~M.}\ \bibnamefont {Horowitz}},\ and\ \bibinfo {author} {\bibfnamefont {N.}~\bibnamefont {Fakhri}},\ }\href {https://doi.org/10.48550/arXiv.2107.05701} {\bibinfo {title} {Scale-dependent irreversibility in living matter}} (\bibinfo {year} {2021})\BibitemShut {NoStop}%
\bibitem [{\citenamefont {Gingrich}\ \emph {et~al.}(2016)\citenamefont {Gingrich}, \citenamefont {Horowitz}, \citenamefont {Perunov},\ and\ \citenamefont {England}}]{gingrich2016dissipation}%
  \BibitemOpen
  \bibfield  {author} {\bibinfo {author} {\bibfnamefont {T.~R.}\ \bibnamefont {Gingrich}}, \bibinfo {author} {\bibfnamefont {J.~M.}\ \bibnamefont {Horowitz}}, \bibinfo {author} {\bibfnamefont {N.}~\bibnamefont {Perunov}},\ and\ \bibinfo {author} {\bibfnamefont {J.~L.}\ \bibnamefont {England}},\ }\href {https://doi.org/10.1103/PhysRevLett.116.120601} {\bibfield  {journal} {\bibinfo  {journal} {Physical Review Letters}\ }\textbf {\bibinfo {volume} {116}},\ \bibinfo {pages} {120601} (\bibinfo {year} {2016})}\BibitemShut {NoStop}%
\bibitem [{\citenamefont {Seifert}(2019)}]{seifert2019stochastic}%
  \BibitemOpen
  \bibfield  {author} {\bibinfo {author} {\bibfnamefont {U.}~\bibnamefont {Seifert}},\ }\href {https://doi.org/10.1146/annurev-conmatphys-031218-013554} {\bibfield  {journal} {\bibinfo  {journal} {Annual Review of Condensed Matter Physics}\ }\textbf {\bibinfo {volume} {10}},\ \bibinfo {pages} {171} (\bibinfo {year} {2019})}\BibitemShut {NoStop}%
\bibitem [{\citenamefont {Bahri}\ \emph {et~al.}(2020)\citenamefont {Bahri}, \citenamefont {Kadmon}, \citenamefont {Pennington}, \citenamefont {Schoenholz}, \citenamefont {Sohl-Dickstein},\ and\ \citenamefont {Ganguli}}]{bahri2020statistical}%
  \BibitemOpen
  \bibfield  {author} {\bibinfo {author} {\bibfnamefont {Y.}~\bibnamefont {Bahri}}, \bibinfo {author} {\bibfnamefont {J.}~\bibnamefont {Kadmon}}, \bibinfo {author} {\bibfnamefont {J.}~\bibnamefont {Pennington}}, \bibinfo {author} {\bibfnamefont {S.~S.}\ \bibnamefont {Schoenholz}}, \bibinfo {author} {\bibfnamefont {J.}~\bibnamefont {Sohl-Dickstein}},\ and\ \bibinfo {author} {\bibfnamefont {S.}~\bibnamefont {Ganguli}},\ }\href {https://doi.org/10.1146/annurev-conmatphys-031119-050745} {\bibfield  {journal} {\bibinfo  {journal} {Annual Review of Condensed Matter Physics}\ }\textbf {\bibinfo {volume} {11}},\ \bibinfo {pages} {501} (\bibinfo {year} {2020})}\BibitemShut {NoStop}%
\bibitem [{\citenamefont {Lusch}\ \emph {et~al.}(2018)\citenamefont {Lusch}, \citenamefont {Kutz},\ and\ \citenamefont {Brunton}}]{lusch2018deep}%
  \BibitemOpen
  \bibfield  {author} {\bibinfo {author} {\bibfnamefont {B.}~\bibnamefont {Lusch}}, \bibinfo {author} {\bibfnamefont {J.~N.}\ \bibnamefont {Kutz}},\ and\ \bibinfo {author} {\bibfnamefont {S.~L.}\ \bibnamefont {Brunton}},\ }\href {https://doi.org/10.1038/s41467-018-07210-0} {\bibfield  {journal} {\bibinfo  {journal} {Nature Communications}\ }\textbf {\bibinfo {volume} {9}},\ \bibinfo {pages} {1} (\bibinfo {year} {2018})}\BibitemShut {NoStop}%
\bibitem [{\citenamefont {Falk}\ \emph {et~al.}(2021)\citenamefont {Falk}, \citenamefont {Alizadehyazdi}, \citenamefont {Jaeger},\ and\ \citenamefont {Murugan}}]{falk2021learning}%
  \BibitemOpen
  \bibfield  {author} {\bibinfo {author} {\bibfnamefont {M.~J.}\ \bibnamefont {Falk}}, \bibinfo {author} {\bibfnamefont {V.}~\bibnamefont {Alizadehyazdi}}, \bibinfo {author} {\bibfnamefont {H.}~\bibnamefont {Jaeger}},\ and\ \bibinfo {author} {\bibfnamefont {A.}~\bibnamefont {Murugan}},\ }\href {https://doi.org/10.1103/PhysRevResearch.3.033291} {\bibfield  {journal} {\bibinfo  {journal} {Physical Review Research}\ }\textbf {\bibinfo {volume} {3}},\ \bibinfo {pages} {033291} (\bibinfo {year} {2021})}\BibitemShut {NoStop}%
\bibitem [{\citenamefont {Schmitt}\ \emph {et~al.}(2023)\citenamefont {Schmitt}, \citenamefont {Colen}, \citenamefont {Sala}, \citenamefont {Devany}, \citenamefont {Seetharaman}, \citenamefont {Gardel}, \citenamefont {Oakes},\ and\ \citenamefont {Vitelli}}]{schmitt2023zyxin}%
  \BibitemOpen
  \bibfield  {author} {\bibinfo {author} {\bibfnamefont {M.~S.}\ \bibnamefont {Schmitt}}, \bibinfo {author} {\bibfnamefont {J.}~\bibnamefont {Colen}}, \bibinfo {author} {\bibfnamefont {S.}~\bibnamefont {Sala}}, \bibinfo {author} {\bibfnamefont {J.}~\bibnamefont {Devany}}, \bibinfo {author} {\bibfnamefont {S.}~\bibnamefont {Seetharaman}}, \bibinfo {author} {\bibfnamefont {M.~L.}\ \bibnamefont {Gardel}}, \bibinfo {author} {\bibfnamefont {P.~W.}\ \bibnamefont {Oakes}},\ and\ \bibinfo {author} {\bibfnamefont {V.}~\bibnamefont {Vitelli}},\ }\href {https://doi.org/10.48550/arXiv.2303.00176} {\bibinfo {title} {Zyxin is all you need: machine learning adherent cell mechanics}} (\bibinfo {year} {2023})\BibitemShut {NoStop}%
\bibitem [{\citenamefont {Hern\'andez}\ \emph {et~al.}(2023)\citenamefont {Hern\'andez}, \citenamefont {Roman},\ and\ \citenamefont {Nemenman}}]{hernandez2023low}%
  \BibitemOpen
  \bibfield  {author} {\bibinfo {author} {\bibfnamefont {D.~G.}\ \bibnamefont {Hern\'andez}}, \bibinfo {author} {\bibfnamefont {A.}~\bibnamefont {Roman}},\ and\ \bibinfo {author} {\bibfnamefont {I.}~\bibnamefont {Nemenman}},\ }\href {https://doi.org/10.1103/PhysRevE.108.014101} {\bibfield  {journal} {\bibinfo  {journal} {Physical Review E}\ }\textbf {\bibinfo {volume} {108}},\ \bibinfo {pages} {014101} (\bibinfo {year} {2023})}\BibitemShut {NoStop}%
\bibitem [{\citenamefont {Kim}\ and\ \citenamefont {Mnih}(2018)}]{kim2018disentangling}%
  \BibitemOpen
  \bibfield  {author} {\bibinfo {author} {\bibfnamefont {H.}~\bibnamefont {Kim}}\ and\ \bibinfo {author} {\bibfnamefont {A.}~\bibnamefont {Mnih}},\ }in\ \href {https://doi.org/10.48550/arXiv.1802.05983} {\emph {\bibinfo {booktitle} {Proceedings of the 35th International Conference on Machine Learning, {ICML} 2018, Stockholmsm{\"{a}}ssan, Stockholm, Sweden, July 10-15, 2018}}},\ \bibinfo {series} {Proceedings of Machine Learning Research}, Vol.~\bibinfo {volume} {80},\ \bibinfo {editor} {edited by\ \bibinfo {editor} {\bibfnamefont {J.~G.}\ \bibnamefont {Dy}}\ and\ \bibinfo {editor} {\bibfnamefont {A.}~\bibnamefont {Krause}}}\ (\bibinfo  {publisher} {PMLR},\ \bibinfo {year} {2018})\ pp.\ \bibinfo {pages} {2649--2658}\BibitemShut {NoStop}%
\bibitem [{\citenamefont {Tan}\ \emph {et~al.}(2020)\citenamefont {Tan}, \citenamefont {Liu}, \citenamefont {Miller}, \citenamefont {Tekant}, \citenamefont {Dunkel},\ and\ \citenamefont {Fakhri}}]{tan2020topological}%
  \BibitemOpen
  \bibfield  {author} {\bibinfo {author} {\bibfnamefont {T.~H.}\ \bibnamefont {Tan}}, \bibinfo {author} {\bibfnamefont {J.}~\bibnamefont {Liu}}, \bibinfo {author} {\bibfnamefont {P.~W.}\ \bibnamefont {Miller}}, \bibinfo {author} {\bibfnamefont {M.}~\bibnamefont {Tekant}}, \bibinfo {author} {\bibfnamefont {J.}~\bibnamefont {Dunkel}},\ and\ \bibinfo {author} {\bibfnamefont {N.}~\bibnamefont {Fakhri}},\ }\href {https://doi.org/10.1038/s41567-020-0841-9} {\bibfield  {journal} {\bibinfo  {journal} {Nature Physics}\ }\textbf {\bibinfo {volume} {16}},\ \bibinfo {pages} {657} (\bibinfo {year} {2020})}\BibitemShut {NoStop}%
\bibitem [{\citenamefont {Wigbers}\ \emph {et~al.}(2021)\citenamefont {Wigbers}, \citenamefont {Tan}, \citenamefont {Brauns}, \citenamefont {Liu}, \citenamefont {Swartz}, \citenamefont {Frey},\ and\ \citenamefont {Fakhri}}]{wigbers2021hierarchy}%
  \BibitemOpen
  \bibfield  {author} {\bibinfo {author} {\bibfnamefont {M.~C.}\ \bibnamefont {Wigbers}}, \bibinfo {author} {\bibfnamefont {T.~H.}\ \bibnamefont {Tan}}, \bibinfo {author} {\bibfnamefont {F.}~\bibnamefont {Brauns}}, \bibinfo {author} {\bibfnamefont {J.}~\bibnamefont {Liu}}, \bibinfo {author} {\bibfnamefont {S.~Z.}\ \bibnamefont {Swartz}}, \bibinfo {author} {\bibfnamefont {E.}~\bibnamefont {Frey}},\ and\ \bibinfo {author} {\bibfnamefont {N.}~\bibnamefont {Fakhri}},\ }\href {https://doi.org/10.1038/s41567-021-01164-9} {\bibfield  {journal} {\bibinfo  {journal} {Nature Physics}\ }\textbf {\bibinfo {volume} {17}},\ \bibinfo {pages} {578} (\bibinfo {year} {2021})}\BibitemShut {NoStop}%
\bibitem [{\citenamefont {Falasco}\ \emph {et~al.}(2018)\citenamefont {Falasco}, \citenamefont {Rao},\ and\ \citenamefont {Esposito}}]{falasco2018information}%
  \BibitemOpen
  \bibfield  {author} {\bibinfo {author} {\bibfnamefont {G.}~\bibnamefont {Falasco}}, \bibinfo {author} {\bibfnamefont {R.}~\bibnamefont {Rao}},\ and\ \bibinfo {author} {\bibfnamefont {M.}~\bibnamefont {Esposito}},\ }\href {https://doi.org/10.1103/PhysRevLett.121.108301} {\bibfield  {journal} {\bibinfo  {journal} {Physical Review Letters}\ }\textbf {\bibinfo {volume} {121}},\ \bibinfo {pages} {108301} (\bibinfo {year} {2018})}\BibitemShut {NoStop}%
\bibitem [{\citenamefont {Kuramoto}(1984)}]{kuramoto1984chemical}%
  \BibitemOpen
  \bibfield  {author} {\bibinfo {author} {\bibfnamefont {Y.}~\bibnamefont {Kuramoto}},\ }\href {https://doi.org/10.1007/978-3-642-69689-3} {\emph {\bibinfo {title} {Chemical Oscillations, Waves, and Turbulence}}}\ (\bibinfo  {publisher} {Springer},\ \bibinfo {year} {1984})\ pp.\ \bibinfo {pages} {111--140}\BibitemShut {NoStop}%
\bibitem [{\citenamefont {Liu}\ \emph {et~al.}(2021)\citenamefont {Liu}, \citenamefont {Totz}, \citenamefont {Miller}, \citenamefont {Hastewell}, \citenamefont {Chao}, \citenamefont {Dunkel},\ and\ \citenamefont {Fakhri}}]{liu2021topological}%
  \BibitemOpen
  \bibfield  {author} {\bibinfo {author} {\bibfnamefont {J.}~\bibnamefont {Liu}}, \bibinfo {author} {\bibfnamefont {J.~F.}\ \bibnamefont {Totz}}, \bibinfo {author} {\bibfnamefont {P.~W.}\ \bibnamefont {Miller}}, \bibinfo {author} {\bibfnamefont {A.~D.}\ \bibnamefont {Hastewell}}, \bibinfo {author} {\bibfnamefont {Y.-C.}\ \bibnamefont {Chao}}, \bibinfo {author} {\bibfnamefont {J.}~\bibnamefont {Dunkel}},\ and\ \bibinfo {author} {\bibfnamefont {N.}~\bibnamefont {Fakhri}},\ }\href {https://doi.org/10.1073/pnas.2104191118} {\bibfield  {journal} {\bibinfo  {journal} {Proceedings of the National Academy of Sciences}\ }\textbf {\bibinfo {volume} {118}},\ \bibinfo {pages} {e2104191118} (\bibinfo {year} {2021})}\BibitemShut {NoStop}%
\bibitem [{\citenamefont {Kingma}\ and\ \citenamefont {Welling}(2014)}]{kingma2013autoencoding}%
  \BibitemOpen
  \bibfield  {author} {\bibinfo {author} {\bibfnamefont {D.~P.}\ \bibnamefont {Kingma}}\ and\ \bibinfo {author} {\bibfnamefont {M.}~\bibnamefont {Welling}},\ }in\ \href {https://doi.org/10.48550/arXiv.1312.6114} {\emph {\bibinfo {booktitle} {2nd International Conference on Learning Representations, {ICLR} 2014, Banff, AB, Canada, April 14-16, 2014, Conference Track Proceedings}}},\ \bibinfo {editor} {edited by\ \bibinfo {editor} {\bibfnamefont {Y.}~\bibnamefont {Bengio}}\ and\ \bibinfo {editor} {\bibfnamefont {Y.}~\bibnamefont {LeCun}}}\ (\bibinfo  {publisher} {arXiv},\ \bibinfo {year} {2014})\BibitemShut {NoStop}%
\bibitem [{\citenamefont {Guyon}\ \emph {et~al.}(1991)\citenamefont {Guyon}, \citenamefont {Albrecht}, \citenamefont {LeCun}, \citenamefont {Denker},\ and\ \citenamefont {Hubbard}}]{guyon1991design}%
  \BibitemOpen
  \bibfield  {author} {\bibinfo {author} {\bibfnamefont {I.}~\bibnamefont {Guyon}}, \bibinfo {author} {\bibfnamefont {P.}~\bibnamefont {Albrecht}}, \bibinfo {author} {\bibfnamefont {Y.}~\bibnamefont {LeCun}}, \bibinfo {author} {\bibfnamefont {J.}~\bibnamefont {Denker}},\ and\ \bibinfo {author} {\bibfnamefont {W.}~\bibnamefont {Hubbard}},\ }\href {https://doi.org/10.1016/0031-3203(91)90081-F} {\bibfield  {journal} {\bibinfo  {journal} {Pattern Recognition}\ }\textbf {\bibinfo {volume} {24}},\ \bibinfo {pages} {105} (\bibinfo {year} {1991})}\BibitemShut {NoStop}%
\bibitem [{\citenamefont {Heffernan}\ \emph {et~al.}(2017)\citenamefont {Heffernan}, \citenamefont {Yang}, \citenamefont {Paliwal},\ and\ \citenamefont {Zhou}}]{heffernan2017capturing}%
  \BibitemOpen
  \bibfield  {author} {\bibinfo {author} {\bibfnamefont {R.}~\bibnamefont {Heffernan}}, \bibinfo {author} {\bibfnamefont {Y.}~\bibnamefont {Yang}}, \bibinfo {author} {\bibfnamefont {K.}~\bibnamefont {Paliwal}},\ and\ \bibinfo {author} {\bibfnamefont {Y.}~\bibnamefont {Zhou}},\ }\href {https://doi.org/10.1093/bioinformatics/btx218} {\bibfield  {journal} {\bibinfo  {journal} {Bioinformatics}\ }\textbf {\bibinfo {volume} {33}},\ \bibinfo {pages} {2842} (\bibinfo {year} {2017})}\BibitemShut {NoStop}%
\bibitem [{\citenamefont {Gabbard}\ \emph {et~al.}(2021)\citenamefont {Gabbard}, \citenamefont {Messenger}, \citenamefont {Heng}, \citenamefont {Tonolini},\ and\ \citenamefont {Murray-Smith}}]{gabbard2021bayesian}%
  \BibitemOpen
  \bibfield  {author} {\bibinfo {author} {\bibfnamefont {H.}~\bibnamefont {Gabbard}}, \bibinfo {author} {\bibfnamefont {C.}~\bibnamefont {Messenger}}, \bibinfo {author} {\bibfnamefont {I.~S.}\ \bibnamefont {Heng}}, \bibinfo {author} {\bibfnamefont {F.}~\bibnamefont {Tonolini}},\ and\ \bibinfo {author} {\bibfnamefont {R.}~\bibnamefont {Murray-Smith}},\ }\href {https://doi.org/10.1038/s41567-021-01425-7} {\bibfield  {journal} {\bibinfo  {journal} {Nature Physics}\ }\textbf {\bibinfo {volume} {18}},\ \bibinfo {pages} {112} (\bibinfo {year} {2021})}\BibitemShut {NoStop}%
\bibitem [{\citenamefont {Miles}\ \emph {et~al.}(2021)\citenamefont {Miles}, \citenamefont {Carbone}, \citenamefont {Sturm}, \citenamefont {Lu}, \citenamefont {Weichselbaum}, \citenamefont {Barros},\ and\ \citenamefont {Konik}}]{miles2021machine}%
  \BibitemOpen
  \bibfield  {author} {\bibinfo {author} {\bibfnamefont {C.}~\bibnamefont {Miles}}, \bibinfo {author} {\bibfnamefont {M.~R.}\ \bibnamefont {Carbone}}, \bibinfo {author} {\bibfnamefont {E.~J.}\ \bibnamefont {Sturm}}, \bibinfo {author} {\bibfnamefont {D.}~\bibnamefont {Lu}}, \bibinfo {author} {\bibfnamefont {A.}~\bibnamefont {Weichselbaum}}, \bibinfo {author} {\bibfnamefont {K.}~\bibnamefont {Barros}},\ and\ \bibinfo {author} {\bibfnamefont {R.~M.}\ \bibnamefont {Konik}},\ }\href {https://doi.org/10.1103/PhysRevB.104.235111} {\bibfield  {journal} {\bibinfo  {journal} {Physical Review B}\ }\textbf {\bibinfo {volume} {104}},\ \bibinfo {pages} {235111} (\bibinfo {year} {2021})}\BibitemShut {NoStop}%
\bibitem [{\citenamefont {Takeishi}\ and\ \citenamefont {Kalousis}(2021)}]{takeishi2021physics}%
  \BibitemOpen
  \bibfield  {author} {\bibinfo {author} {\bibfnamefont {N.}~\bibnamefont {Takeishi}}\ and\ \bibinfo {author} {\bibfnamefont {A.}~\bibnamefont {Kalousis}},\ }in\ \href {https://doi.org/10.48550/arXiv.2102.13156} {\emph {\bibinfo {booktitle} {Advances in Neural Information Processing Systems 34: Annual Conference on Neural Information Processing Systems 2021, NeurIPS 2021, December 6-14, 2021, virtual}}},\ \bibinfo {editor} {edited by\ \bibinfo {editor} {\bibfnamefont {M.}~\bibnamefont {Ranzato}}, \bibinfo {editor} {\bibfnamefont {A.}~\bibnamefont {Beygelzimer}}, \bibinfo {editor} {\bibfnamefont {Y.~N.}\ \bibnamefont {Dauphin}}, \bibinfo {editor} {\bibfnamefont {P.}~\bibnamefont {Liang}},\ and\ \bibinfo {editor} {\bibfnamefont {J.}~\bibnamefont {Wortman~Vaughan}}}\ (\bibinfo  {publisher} {Curran Associates Inc., Red Hook, NY, United States},\ \bibinfo {year} {2021})\ pp.\ \bibinfo {pages} {14809--14821}\BibitemShut {NoStop}%
\bibitem [{\citenamefont {Wang}\ \emph {et~al.}(2021)\citenamefont {Wang}, \citenamefont {He}, \citenamefont {Li}, \citenamefont {Chen}, \citenamefont {Zhai},\ and\ \citenamefont {Zhang}}]{wang2021flow}%
  \BibitemOpen
  \bibfield  {author} {\bibinfo {author} {\bibfnamefont {J.}~\bibnamefont {Wang}}, \bibinfo {author} {\bibfnamefont {C.}~\bibnamefont {He}}, \bibinfo {author} {\bibfnamefont {R.}~\bibnamefont {Li}}, \bibinfo {author} {\bibfnamefont {H.}~\bibnamefont {Chen}}, \bibinfo {author} {\bibfnamefont {C.}~\bibnamefont {Zhai}},\ and\ \bibinfo {author} {\bibfnamefont {M.}~\bibnamefont {Zhang}},\ }\href {https://doi.org/10.1063/5.0053979} {\bibfield  {journal} {\bibinfo  {journal} {Physics of Fluids}\ }\textbf {\bibinfo {volume} {33}},\ \bibinfo {pages} {086108} (\bibinfo {year} {2021})}\BibitemShut {NoStop}%
\bibitem [{\citenamefont {Im}\ \emph {et~al.}(2017)\citenamefont {Im}, \citenamefont {Ahn}, \citenamefont {Memisevic},\ and\ \citenamefont {Bengio}}]{im2017denoising}%
  \BibitemOpen
  \bibfield  {author} {\bibinfo {author} {\bibfnamefont {D.~I.}\ \bibnamefont {Im}}, \bibinfo {author} {\bibfnamefont {S.}~\bibnamefont {Ahn}}, \bibinfo {author} {\bibfnamefont {R.}~\bibnamefont {Memisevic}},\ and\ \bibinfo {author} {\bibfnamefont {Y.}~\bibnamefont {Bengio}},\ }in\ \href {https://doi.org/10.1609/aaai.v31i1.10777} {\emph {\bibinfo {booktitle} {Proceedings of the AAAI Conference on Artificial Intelligence}}},\ Vol.~\bibinfo {volume} {31}\ (\bibinfo  {publisher} {AAAI Press, San Francisco, CA},\ \bibinfo {year} {2017})\ pp.\ \bibinfo {pages} {2059--2065}\BibitemShut {NoStop}%
\bibitem [{\citenamefont {Liu}\ \emph {et~al.}(2020)\citenamefont {Liu}, \citenamefont {Siu}, \citenamefont {Wang}, \citenamefont {Li},\ and\ \citenamefont {Cani}}]{liu2020unsupervised}%
  \BibitemOpen
  \bibfield  {author} {\bibinfo {author} {\bibfnamefont {Z.-S.}\ \bibnamefont {Liu}}, \bibinfo {author} {\bibfnamefont {W.-C.}\ \bibnamefont {Siu}}, \bibinfo {author} {\bibfnamefont {L.-W.}\ \bibnamefont {Wang}}, \bibinfo {author} {\bibfnamefont {C.-T.}\ \bibnamefont {Li}},\ and\ \bibinfo {author} {\bibfnamefont {M.-P.}\ \bibnamefont {Cani}},\ }in\ \href {https://doi.org/https://doi.org/10.1109/CVPRW50498.2020.00229} {\emph {\bibinfo {booktitle} {Proceedings of the IEEE/CVF Conference on Computer Vision and Pattern Recognition (CVPR) Workshops}}}\ (\bibinfo  {publisher} {IEEE},\ \bibinfo {year} {2020})\ pp.\ \bibinfo {pages} {1788--1797}\BibitemShut {NoStop}%
\bibitem [{\citenamefont {Ziv}\ and\ \citenamefont {Lempel}(1977)}]{ziv1977universal}%
  \BibitemOpen
  \bibfield  {author} {\bibinfo {author} {\bibfnamefont {J.}~\bibnamefont {Ziv}}\ and\ \bibinfo {author} {\bibfnamefont {A.}~\bibnamefont {Lempel}},\ }\href {https://doi.org/10.1109/TIT.1977.1055714} {\bibfield  {journal} {\bibinfo  {journal} {IEEE Transactions on Information Theory}\ }\textbf {\bibinfo {volume} {23}},\ \bibinfo {pages} {337} (\bibinfo {year} {1977})}\BibitemShut {NoStop}%
\bibitem [{\citenamefont {Rold{\'a}n}\ and\ \citenamefont {Parrondo}(2010)}]{roldan2010estimating}%
  \BibitemOpen
  \bibfield  {author} {\bibinfo {author} {\bibfnamefont {{\'E}.}~\bibnamefont {Rold{\'a}n}}\ and\ \bibinfo {author} {\bibfnamefont {J.~M.~R.}\ \bibnamefont {Parrondo}},\ }\href {https://doi.org/10.1103/PhysRevLett.105.150607} {\bibfield  {journal} {\bibinfo  {journal} {Physical Review Letters}\ }\textbf {\bibinfo {volume} {105}},\ \bibinfo {pages} {150607} (\bibinfo {year} {2010})}\BibitemShut {NoStop}%
\bibitem [{\citenamefont {Rold{\'a}n}\ and\ \citenamefont {Parrondo}(2012)}]{roldan2012entropy}%
  \BibitemOpen
  \bibfield  {author} {\bibinfo {author} {\bibfnamefont {{\'E}.}~\bibnamefont {Rold{\'a}n}}\ and\ \bibinfo {author} {\bibfnamefont {J.~M.~R.}\ \bibnamefont {Parrondo}},\ }\href {https://doi.org/10.1103/PhysRevE.85.031129} {\bibfield  {journal} {\bibinfo  {journal} {Physical Review E}\ }\textbf {\bibinfo {volume} {85}},\ \bibinfo {pages} {031129} (\bibinfo {year} {2012})}\BibitemShut {NoStop}%
\bibitem [{\citenamefont {Johnson}\ and\ \citenamefont {Lindenstrauss}(1984)}]{johnson1984extensions}%
  \BibitemOpen
  \bibfield  {author} {\bibinfo {author} {\bibfnamefont {W.~B.}\ \bibnamefont {Johnson}}\ and\ \bibinfo {author} {\bibfnamefont {J.}~\bibnamefont {Lindenstrauss}},\ }in\ \href {https://doi.org/10.1090/conm/026/737400} {\emph {\bibinfo {booktitle} {Conference in Modern Analysis and Probability}}},\ \bibinfo {series} {Contemporary Mathematics}, Vol.~\bibinfo {volume} {26}\ (\bibinfo  {publisher} {American Mathematical Society},\ \bibinfo {year} {1984})\ pp.\ \bibinfo {pages} {189--206}\BibitemShut {NoStop}%
\bibitem [{\citenamefont {Jordan}\ and\ \citenamefont {Dimakis}(2021)}]{jordan2021provable}%
  \BibitemOpen
  \bibfield  {author} {\bibinfo {author} {\bibfnamefont {M.}~\bibnamefont {Jordan}}\ and\ \bibinfo {author} {\bibfnamefont {A.~G.}\ \bibnamefont {Dimakis}},\ }in\ \href {https://doi.org/10.48550/arXiv.2107.02732} {\emph {\bibinfo {booktitle} {Proceedings of the 38th International Conference on Machine Learning, {ICML} 2021, 18-24 July 2021, Virtual Event}}},\ \bibinfo {series} {Proceedings of Machine Learning Research}, Vol.\ \bibinfo {volume} {139},\ \bibinfo {editor} {edited by\ \bibinfo {editor} {\bibfnamefont {M.}~\bibnamefont {Meila}}\ and\ \bibinfo {editor} {\bibfnamefont {T.}~\bibnamefont {Zhang}}}\ (\bibinfo  {publisher} {PMLR},\ \bibinfo {year} {2021})\ pp.\ \bibinfo {pages} {5118--5126}\BibitemShut {NoStop}%
\bibitem [{\citenamefont {Camuto}\ and\ \citenamefont {Willetts}(2022)}]{camuto2022variational}%
  \BibitemOpen
  \bibfield  {author} {\bibinfo {author} {\bibfnamefont {A.}~\bibnamefont {Camuto}}\ and\ \bibinfo {author} {\bibfnamefont {M.}~\bibnamefont {Willetts}},\ }in\ \href {https://doi.org/10.48550/arXiv.2105.14866} {\emph {\bibinfo {booktitle} {International Conference on Artificial Intelligence and Statistics, {AISTATS} 2022, 28-30 March 2022, Virtual Event}}},\ \bibinfo {series} {Proceedings of Machine Learning Research}, Vol.\ \bibinfo {volume} {151}\ (\bibinfo  {publisher} {PMLR},\ \bibinfo {year} {2022})\ pp.\ \bibinfo {pages} {4595--4611}\BibitemShut {NoStop}%
\bibitem [{\citenamefont {Ro}\ \emph {et~al.}(2022)\citenamefont {Ro}, \citenamefont {Guo}, \citenamefont {Shih}, \citenamefont {Phan}, \citenamefont {Austin}, \citenamefont {Levine}, \citenamefont {Chaikin},\ and\ \citenamefont {Martiniani}}]{ro2022model}%
  \BibitemOpen
  \bibfield  {author} {\bibinfo {author} {\bibfnamefont {S.}~\bibnamefont {Ro}}, \bibinfo {author} {\bibfnamefont {B.}~\bibnamefont {Guo}}, \bibinfo {author} {\bibfnamefont {A.}~\bibnamefont {Shih}}, \bibinfo {author} {\bibfnamefont {T.~V.}\ \bibnamefont {Phan}}, \bibinfo {author} {\bibfnamefont {R.~H.}\ \bibnamefont {Austin}}, \bibinfo {author} {\bibfnamefont {D.}~\bibnamefont {Levine}}, \bibinfo {author} {\bibfnamefont {P.~M.}\ \bibnamefont {Chaikin}},\ and\ \bibinfo {author} {\bibfnamefont {S.}~\bibnamefont {Martiniani}},\ }\href {https://doi.org/10.1103/PhysRevLett.129.220601} {\bibfield  {journal} {\bibinfo  {journal} {Phys. Rev. Lett.}\ }\textbf {\bibinfo {volume} {129}},\ \bibinfo {pages} {220601} (\bibinfo {year} {2022})}\BibitemShut {NoStop}%
\bibitem [{\citenamefont {Pelling}\ \emph {et~al.}(2004)\citenamefont {Pelling}, \citenamefont {Sehati}, \citenamefont {Gralla}, \citenamefont {Valentine},\ and\ \citenamefont {Gimzewksi}}]{pelling2004local}%
  \BibitemOpen
  \bibfield  {author} {\bibinfo {author} {\bibfnamefont {A.~E.}\ \bibnamefont {Pelling}}, \bibinfo {author} {\bibfnamefont {S.}~\bibnamefont {Sehati}}, \bibinfo {author} {\bibfnamefont {E.~B.}\ \bibnamefont {Gralla}}, \bibinfo {author} {\bibfnamefont {J.~S.}\ \bibnamefont {Valentine}},\ and\ \bibinfo {author} {\bibfnamefont {J.~K.}\ \bibnamefont {Gimzewksi}},\ }\href {https://doi.org/10.1126/science.1097640} {\bibfield  {journal} {\bibinfo  {journal} {Science}\ }\textbf {\bibinfo {volume} {305}},\ \bibinfo {pages} {1147} (\bibinfo {year} {2004})}\BibitemShut {NoStop}%
\bibitem [{\citenamefont {Aranson}\ and\ \citenamefont {Kramer}(2002)}]{aranson2002world}%
  \BibitemOpen
  \bibfield  {author} {\bibinfo {author} {\bibfnamefont {I.~S.}\ \bibnamefont {Aranson}}\ and\ \bibinfo {author} {\bibfnamefont {L.}~\bibnamefont {Kramer}},\ }\href {https://doi.org/10.1103/RevModPhys.74.99} {\bibfield  {journal} {\bibinfo  {journal} {Reviews of Modern Physics}\ }\textbf {\bibinfo {volume} {74}},\ \bibinfo {pages} {99} (\bibinfo {year} {2002})}\BibitemShut {NoStop}%
\bibitem [{\citenamefont {Chat{\'e}}\ and\ \citenamefont {Manneville}(1996)}]{chate1996phase}%
  \BibitemOpen
  \bibfield  {author} {\bibinfo {author} {\bibfnamefont {H.}~\bibnamefont {Chat{\'e}}}\ and\ \bibinfo {author} {\bibfnamefont {P.}~\bibnamefont {Manneville}},\ }\href {https://doi.org/10.1016/0378-4371(95)00361-4} {\bibfield  {journal} {\bibinfo  {journal} {Physica A: Statistical Mechanics and its Applications}\ }\textbf {\bibinfo {volume} {224}},\ \bibinfo {pages} {348} (\bibinfo {year} {1996})}\BibitemShut {NoStop}%
\bibitem [{\citenamefont {Lynn}\ \emph {et~al.}(2022)\citenamefont {Lynn}, \citenamefont {Holmes}, \citenamefont {Bialek},\ and\ \citenamefont {Schwab}}]{lynn2022decomposing}%
  \BibitemOpen
  \bibfield  {author} {\bibinfo {author} {\bibfnamefont {C.~W.}\ \bibnamefont {Lynn}}, \bibinfo {author} {\bibfnamefont {C.~M.}\ \bibnamefont {Holmes}}, \bibinfo {author} {\bibfnamefont {W.}~\bibnamefont {Bialek}},\ and\ \bibinfo {author} {\bibfnamefont {D.~J.}\ \bibnamefont {Schwab}},\ }\href {https://doi.org/10.1103/PhysRevLett.129.118101} {\bibfield  {journal} {\bibinfo  {journal} {Physical Review Letters}\ }\textbf {\bibinfo {volume} {129}},\ \bibinfo {pages} {118101} (\bibinfo {year} {2022})}\BibitemShut {NoStop}%
\bibitem [{\citenamefont {Skinner}\ and\ \citenamefont {Dunkel}(2021)}]{skinner2021estimating}%
  \BibitemOpen
  \bibfield  {author} {\bibinfo {author} {\bibfnamefont {D.~J.}\ \bibnamefont {Skinner}}\ and\ \bibinfo {author} {\bibfnamefont {J.}~\bibnamefont {Dunkel}},\ }\href {https://doi.org/10.1103/PhysRevLett.127.198101} {\bibfield  {journal} {\bibinfo  {journal} {Physical Review Letters}\ }\textbf {\bibinfo {volume} {127}},\ \bibinfo {pages} {198101} (\bibinfo {year} {2021})}\BibitemShut {NoStop}%
\bibitem [{\citenamefont {Yu}\ \emph {et~al.}(2021)\citenamefont {Yu}, \citenamefont {Zhang},\ and\ \citenamefont {Tu}}]{yu2021inverse}%
  \BibitemOpen
  \bibfield  {author} {\bibinfo {author} {\bibfnamefont {Q.}~\bibnamefont {Yu}}, \bibinfo {author} {\bibfnamefont {D.}~\bibnamefont {Zhang}},\ and\ \bibinfo {author} {\bibfnamefont {Y.}~\bibnamefont {Tu}},\ }\href {https://doi.org/10.1103/PhysRevLett.126.080601} {\bibfield  {journal} {\bibinfo  {journal} {Physical Review Letters}\ }\textbf {\bibinfo {volume} {126}},\ \bibinfo {pages} {080601} (\bibinfo {year} {2021})}\BibitemShut {NoStop}%
\bibitem [{\citenamefont {Watanabe}(1960)}]{watanabe1960information}%
  \BibitemOpen
  \bibfield  {author} {\bibinfo {author} {\bibfnamefont {S.}~\bibnamefont {Watanabe}},\ }\href {https://doi.org/10.1147/rd.41.0066} {\bibfield  {journal} {\bibinfo  {journal} {IBM Journal of Research and Development}\ }\textbf {\bibinfo {volume} {4}},\ \bibinfo {pages} {66} (\bibinfo {year} {1960})}\BibitemShut {NoStop}%
\bibitem [{\citenamefont {Arcones}\ and\ \citenamefont {Gine}(1992)}]{arcones1992bootstrap}%
  \BibitemOpen
  \bibfield  {author} {\bibinfo {author} {\bibfnamefont {M.~A.}\ \bibnamefont {Arcones}}\ and\ \bibinfo {author} {\bibfnamefont {E.}~\bibnamefont {Gine}},\ }\href {https://doi.org/10.1214/aos/1176348650} {\bibfield  {journal} {\bibinfo  {journal} {Annals of Statistics}\ }\textbf {\bibinfo {volume} {20}},\ \bibinfo {pages} {655} (\bibinfo {year} {1992})}\BibitemShut {NoStop}%
\bibitem [{\citenamefont {Nguyen}\ \emph {et~al.}(2010)\citenamefont {Nguyen}, \citenamefont {Wainwright},\ and\ \citenamefont {Jordan}}]{nguyen2010estimating}%
  \BibitemOpen
  \bibfield  {author} {\bibinfo {author} {\bibfnamefont {X.}~\bibnamefont {Nguyen}}, \bibinfo {author} {\bibfnamefont {M.~J.}\ \bibnamefont {Wainwright}},\ and\ \bibinfo {author} {\bibfnamefont {M.~I.}\ \bibnamefont {Jordan}},\ }\href {https://doi.org/10.1109/TIT.2010.2068870} {\bibfield  {journal} {\bibinfo  {journal} {IEEE Transactions on Information Theory}\ }\textbf {\bibinfo {volume} {56}},\ \bibinfo {pages} {5847} (\bibinfo {year} {2010})}\BibitemShut {NoStop}%
\bibitem [{\citenamefont {Sugiyama}\ \emph {et~al.}(2012)\citenamefont {Sugiyama}, \citenamefont {Suzuki},\ and\ \citenamefont {Kanamori}}]{sugiyama2012density}%
  \BibitemOpen
  \bibfield  {author} {\bibinfo {author} {\bibfnamefont {M.}~\bibnamefont {Sugiyama}}, \bibinfo {author} {\bibfnamefont {T.}~\bibnamefont {Suzuki}},\ and\ \bibinfo {author} {\bibfnamefont {T.}~\bibnamefont {Kanamori}},\ }\href {https://doi.org/10.1007/s10463-011-0343-8} {\bibfield  {journal} {\bibinfo  {journal} {Annals of the Institute of Statistical Mathematics}\ }\textbf {\bibinfo {volume} {64}},\ \bibinfo {pages} {1009} (\bibinfo {year} {2012})}\BibitemShut {NoStop}%
\bibitem [{\citenamefont {Dubois}\ \emph {et~al.}(2021)\citenamefont {Dubois}, \citenamefont {Kastanos}, \citenamefont {Lines}, \citenamefont {Melman},\ and\ \citenamefont {Eraslan}}]{dubois2021disentangling}%
  \BibitemOpen
  \bibfield  {author} {\bibinfo {author} {\bibfnamefont {Y.}~\bibnamefont {Dubois}}, \bibinfo {author} {\bibfnamefont {A.}~\bibnamefont {Kastanos}}, \bibinfo {author} {\bibfnamefont {D.}~\bibnamefont {Lines}}, \bibinfo {author} {\bibfnamefont {B.}~\bibnamefont {Melman}},\ and\ \bibinfo {author} {\bibfnamefont {G.}~\bibnamefont {Eraslan}},\ }\href@noop {} {\bibinfo {title} {Disentangled {VAE}}},\ \bibinfo {howpublished} {\url{https://github.com/YannDubs/disentangling-vae}} (\bibinfo {year} {2021})\BibitemShut {NoStop}%
\bibitem [{\citenamefont {Dupont}(2018)}]{dupont2018learning}%
  \BibitemOpen
  \bibfield  {author} {\bibinfo {author} {\bibfnamefont {E.}~\bibnamefont {Dupont}},\ }in\ \href {https://doi.org/10.48550/arXiv.1804.00104} {\emph {\bibinfo {booktitle} {Advances in Neural Information Processing Systems 31: Annual Conference on Neural Information Processing Systems 2018, NeurIPS 2018, December 3-8, 2018, Montr{\'{e}}al, QC, Canada}}},\ \bibinfo {editor} {edited by\ \bibinfo {editor} {\bibfnamefont {S.}~\bibnamefont {Bengio}}, \bibinfo {editor} {\bibfnamefont {H.~M.}\ \bibnamefont {Wallach}}, \bibinfo {editor} {\bibfnamefont {H.}~\bibnamefont {Larochelle}}, \bibinfo {editor} {\bibfnamefont {K.}~\bibnamefont {Grauman}}, \bibinfo {editor} {\bibfnamefont {N.}~\bibnamefont {Cesa{-}Bianchi}},\ and\ \bibinfo {editor} {\bibfnamefont {G.}~\bibnamefont {Roman}}}\ (\bibinfo  {publisher} {Curran Associates Inc., Red Hook, NY, United States},\ \bibinfo {year} {2018})\ pp.\ \bibinfo {pages} {708--718}\BibitemShut {NoStop}%
\bibitem [{\citenamefont {Paszke}\ \emph {et~al.}(2019)\citenamefont {Paszke}, \citenamefont {Gross}, \citenamefont {Massa}, \citenamefont {Lerer}, \citenamefont {Bradbury}, \citenamefont {Chanan}, \citenamefont {Killeen}, \citenamefont {Lin}, \citenamefont {Gimelshein}, \citenamefont {Antiga}, \citenamefont {Desmaison}, \citenamefont {Kopf}, \citenamefont {Yang}, \citenamefont {DeVito}, \citenamefont {Raison}, \citenamefont {Tejani}, \citenamefont {Chilamkurthy}, \citenamefont {Steiner}, \citenamefont {Fang}, \citenamefont {Bai},\ and\ \citenamefont {Chintala}}]{paszke2019pytorch}%
  \BibitemOpen
  \bibfield  {author} {\bibinfo {author} {\bibfnamefont {A.}~\bibnamefont {Paszke}}, \bibinfo {author} {\bibfnamefont {S.}~\bibnamefont {Gross}}, \bibinfo {author} {\bibfnamefont {F.}~\bibnamefont {Massa}}, \bibinfo {author} {\bibfnamefont {A.}~\bibnamefont {Lerer}}, \bibinfo {author} {\bibfnamefont {J.}~\bibnamefont {Bradbury}}, \bibinfo {author} {\bibfnamefont {G.}~\bibnamefont {Chanan}}, \bibinfo {author} {\bibfnamefont {T.}~\bibnamefont {Killeen}}, \bibinfo {author} {\bibfnamefont {Z.}~\bibnamefont {Lin}}, \bibinfo {author} {\bibfnamefont {N.}~\bibnamefont {Gimelshein}}, \bibinfo {author} {\bibfnamefont {L.}~\bibnamefont {Antiga}}, \bibinfo {author} {\bibfnamefont {A.}~\bibnamefont {Desmaison}}, \bibinfo {author} {\bibfnamefont {A.}~\bibnamefont {Kopf}}, \bibinfo {author} {\bibfnamefont {E.}~\bibnamefont {Yang}}, \bibinfo {author} {\bibfnamefont {Z.}~\bibnamefont {DeVito}}, \bibinfo {author} {\bibfnamefont {M.}~\bibnamefont {Raison}}, \bibinfo {author} {\bibfnamefont {A.}~\bibnamefont {Tejani}}, \bibinfo
  {author} {\bibfnamefont {S.}~\bibnamefont {Chilamkurthy}}, \bibinfo {author} {\bibfnamefont {B.}~\bibnamefont {Steiner}}, \bibinfo {author} {\bibfnamefont {L.}~\bibnamefont {Fang}}, \bibinfo {author} {\bibfnamefont {J.}~\bibnamefont {Bai}},\ and\ \bibinfo {author} {\bibfnamefont {S.}~\bibnamefont {Chintala}},\ }in\ \href {https://doi.org/10.48550/arXiv.1912.01703} {\emph {\bibinfo {booktitle} {Advances in Neural Information Processing Systems 32: Annual Conference on Neural Information Processing Systems 2019, NeurIPS 2019, December 8-14, 2019, Vancouver, BC, Canada}}},\ \bibinfo {editor} {edited by\ \bibinfo {editor} {\bibfnamefont {H.~M.}\ \bibnamefont {Wallach}}, \bibinfo {editor} {\bibfnamefont {H.}~\bibnamefont {Larochelle}}, \bibinfo {editor} {\bibfnamefont {A.}~\bibnamefont {Beygelzimer}}, \bibinfo {editor} {\bibfnamefont {F.}~\bibnamefont {d'Alch{\'{e}}{-}Buc}}, \bibinfo {editor} {\bibfnamefont {E.~B.}\ \bibnamefont {Fox}},\ and\ \bibinfo {editor} {\bibfnamefont {R.}~\bibnamefont {Garnett}}}\
  (\bibinfo  {publisher} {Curran Associates Inc., Red Hook, NY, United States},\ \bibinfo {year} {2019})\ pp.\ \bibinfo {pages} {8024--8035}\BibitemShut {NoStop}%
\bibitem [{\citenamefont {Burgess}\ \emph {et~al.}(2018)\citenamefont {Burgess}, \citenamefont {Higgins}, \citenamefont {Pal}, \citenamefont {Matthey}, \citenamefont {Watters}, \citenamefont {Desjardins},\ and\ \citenamefont {Lerchner}}]{burgess2018understanding}%
  \BibitemOpen
  \bibfield  {author} {\bibinfo {author} {\bibfnamefont {C.~P.}\ \bibnamefont {Burgess}}, \bibinfo {author} {\bibfnamefont {I.}~\bibnamefont {Higgins}}, \bibinfo {author} {\bibfnamefont {A.}~\bibnamefont {Pal}}, \bibinfo {author} {\bibfnamefont {L.}~\bibnamefont {Matthey}}, \bibinfo {author} {\bibfnamefont {N.}~\bibnamefont {Watters}}, \bibinfo {author} {\bibfnamefont {G.}~\bibnamefont {Desjardins}},\ and\ \bibinfo {author} {\bibfnamefont {A.}~\bibnamefont {Lerchner}},\ }in\ \href {https://doi.org/10.48550/arXiv.1804.03599} {\emph {\bibinfo {booktitle} {Proceedings of the 2017 NIPS Workshop on Learning Disentangled Representations, 9 December 2017, Long Beach, California, {USA}}}}\ (\bibinfo  {publisher} {arXiv},\ \bibinfo {year} {2018})\BibitemShut {NoStop}%
\bibitem [{\citenamefont {Goodfellow}\ \emph {et~al.}(2016)\citenamefont {Goodfellow}, \citenamefont {Bengio},\ and\ \citenamefont {Courville}}]{goodfellow2016deep}%
  \BibitemOpen
  \bibfield  {author} {\bibinfo {author} {\bibfnamefont {I.}~\bibnamefont {Goodfellow}}, \bibinfo {author} {\bibfnamefont {Y.}~\bibnamefont {Bengio}},\ and\ \bibinfo {author} {\bibfnamefont {A.}~\bibnamefont {Courville}},\ }\href {http://www.deeplearningbook.org} {\emph {\bibinfo {title} {Deep Learning}}}\ (\bibinfo  {publisher} {MIT Press},\ \bibinfo {address} {Cambridge, MA},\ \bibinfo {year} {2016})\BibitemShut {NoStop}%
\bibitem [{\citenamefont {Kingma}\ and\ \citenamefont {Ba}(2015)}]{kingma2015adam}%
  \BibitemOpen
  \bibfield  {author} {\bibinfo {author} {\bibfnamefont {D.}~\bibnamefont {Kingma}}\ and\ \bibinfo {author} {\bibfnamefont {J.~L.}\ \bibnamefont {Ba}},\ }in\ \href {https://doi.org/10.48550/arXiv.1412.6980} {\emph {\bibinfo {booktitle} {3rd International Conference on Learning Representations, {ICLR} 2015, San Diego, CA, USA, May 7-9, 2015, Conference Track Proceedings}}},\ \bibinfo {editor} {edited by\ \bibinfo {editor} {\bibfnamefont {Y.}~\bibnamefont {Bengio}}\ and\ \bibinfo {editor} {\bibfnamefont {Y.}~\bibnamefont {LeCun}}}\ (\bibinfo {year} {2015})\BibitemShut {NoStop}%
\bibitem [{\citenamefont {Cox}\ and\ \citenamefont {Matthews}(2002)}]{cox2002exponential}%
  \BibitemOpen
  \bibfield  {author} {\bibinfo {author} {\bibfnamefont {S.}~\bibnamefont {Cox}}\ and\ \bibinfo {author} {\bibfnamefont {P.}~\bibnamefont {Matthews}},\ }\href {https://doi.org/https://doi.org/10.1006/jcph.2002.6995} {\bibfield  {journal} {\bibinfo  {journal} {Journal of Computational Physics}\ }\textbf {\bibinfo {volume} {176}},\ \bibinfo {pages} {430} (\bibinfo {year} {2002})}\BibitemShut {NoStop}%
\bibitem [{\citenamefont {Winterbottom}(2005)}]{winterbottom2005complex}%
  \BibitemOpen
  \bibfield  {author} {\bibinfo {author} {\bibfnamefont {D.~M.}\ \bibnamefont {Winterbottom}},\ }\href@noop {} {\bibinfo {title} {The complex {G}inzburg-{L}andau equation}},\ \bibinfo {howpublished} {\url{https://github.com/codeinthehole/codeinthehole.com/blob/58ad3d28ddefb64350ec883b291d4dbe1df096f7/www/static/tutorial/files/CGLsim2D.m}} (\bibinfo {year} {2005})\BibitemShut {NoStop}%
\end{thebibliography}%
\clearpage

\onecolumngrid
\beginsupplement
\setcounter{page}{1}

\section*{Supplemental Material}
\begin{table}[h!]
\caption{{\bf Default tuned hyperparameter values.}}
\label{tab:hyperparameters}
\begin{ruledtabular}
\begin{tabular}{ccc}
Hyperparameter & Simulation models & Experiment models\\
\hline
Regression window (window) & $16$ epochs & $32$ epochs\\
Regression stop (stop) & $16$ epochs & $32$ epochs\\
VAE learning rate (vaelr) & $10^{-3}$ & $10^{-3}$\\
Discriminator learning rate (dlr) & $10^{-4}$ & $10^{-4}$\\
\end{tabular}
\end{ruledtabular}
\end{table}
\clearpage

\clearpage

\clearpage

\clearpage

\clearpage

\clearpage

\clearpage

\clearpage

\end{document}